\documentclass[]{aastex63}
\usepackage{graphicx}
\usepackage{bm}
\usepackage{color}
\usepackage{soul}
\usepackage{hyperref}
\usepackage{enumerate}

\newcommand{\beq}{\begin{equation}}
\newcommand{\eeq}{\end{equation}}
\newcommand{\beqn}{\begin{eqnarray}}
\newcommand{\eeqn}{\end{eqnarray}}

\newcommand{\GR}{\rho}

\newcommand{\maxtov}{M_{\rm max}^{\rm sph}}

\begin{document}
\title{Multimessenger Binary Mergers Containing
  Neutron Stars: Gravitational Waves, Jets, and $\boldsymbol{\gamma}$-Ray Bursts}
\author{Milton Ruiz}
\affiliation{Department of Physics, University of Illinois at
  Urbana-Champaign, Urbana, IL 61801}
\author{Stuart L. Shapiro}
\affiliation{Department of Physics, University of Illinois at
  Urbana-Champaign, Urbana, IL 61801}
\affiliation{Department of Astronomy \& NCSA, University of
  Illinois at Urbana-Champaign, Urbana, IL 61801}
\author{Antonios Tsokaros}
\affiliation{Department of Physics, University of Illinois at
  Urbana-Champaign, Urbana, IL 61801}    
\date{\today}

%%%%%%%%%%%%%%%%%%%%
%%%   ABSTRACT   %%%
%%%%%%%%%%%%%%%%%%%%
\begin{abstract}
  Neutron stars (NSs) are extraordinary not only because they are the densest form
  of matter in the visible Universe but also because they can generate magnetic
  fields ten orders of magnitude larger than those currently constructed on Earth.
  The combination of extreme gravity with the enormous electromagnetic (EM) fields
  gives rise to spectacular phenomena like those observed on August 2017 with the
  merger of a binary neutron star system, an event that generated a gravitational
  wave (GW) signal, a short $\gamma$-ray burst (sGRB), and a kilonova. This event
  serves as the highlight so far of the era of multimessenger astronomy.  In this
  review, we present the current state of our theoretical understanding of compact
  binary mergers containing NSs as gleaned from the latest general relativistic 
  magnetohydrodynamic simulations. Such mergers can lead to events like the one
  on August 2017, GW170817, and its EM counterparts, GRB 170817 and AT 2017gfo. In
  addition to exploring the GW emission from binary black hole-neutron star and
  neutron star-neutron star mergers, we also focus on their counterpart EM signals.
  In particular, we are interested in identifying the conditions under which a
  relativistic jet can be launched following these mergers. Such a jet is an
  essential feature of most sGRB models and provides the main conduit  of energy
  from  the central object to the outer radiation regions. Jet properties, including
  their lifetimes and Poynting luminosities, the effects of the initial  magnetic
  field geometries and spins of the coalescing NSs and black holes, as well as their governing
  equation of state, are discussed. Lastly, we present our current understanding
  of how the Blandford-Znajek mechanism arises from merger remnants
  as the trigger for launching jets, if, when and how 
  a horizon is necessary
  for this mechanism, and the possibility that it can turn on in magnetized neutron
  ergostars, which contain ergoregions, but no horizons. 
\end{abstract}

%\maketitle
\keywords{black holes, neutron stars, gravitational waves, short gamma-ray
  bursts, numerical relativity}

%%%%%%%%%%%%%%%%%%%%%%
%%%  Introduction  %%%
%%%%%%%%%%%%%%%%%%%%%%

\section{\bf Introduction}
\label{sec:introduction}
Gravitational wave astronomy was launched in 2015 with the first-ever gravitational
wave (GW) detection of the inspiral and merger of a binary black hole
(BHBH) system as reported by the LIGO/Virgo (LV) scientific collaboration --event GW150914
\citep{LIGO_first_direct_GW,Abbott:2016nmj}. Two years later the simultaneous detection of GWs from an inspiraling
binary neutron star (NSNS) system, event GW170817, and its postmerger emission of
electromagnetic (EM) radiation spurred the era of multimessenger astronomy
\citep{TheLIGOScientific:2017qsa,2017GCN.21517....1K,GBM:2017lvd,Monitor:2017mdv,Abbott:2017wuw}.
Although at present the LV scientific collaboration almost weekly announces new GW signals whose
progenitors may be BHBHs, NSNSs, or black hole-neutron star (BHNS) systems there has been
no robust discovery of a BHNS system yet, while the subsequent NSNS candidates have been
EM ``orphans'' i.e. no EM radiation has been associated with the GWs produced by them.
Merging NSNSs and BHNSs are not only important sources of gravitational radiation, but
also promising candidates for coincident EM counterparts, which could give new insight
into their sources. Namely, GWs are sensitive to the density profile of NSs and their
measurement enforces tight constraints on the equation of state (EOS) that governs matter
at supranuclear densities~\citep{2016PhR...621..127L}, while postmerger EM signatures can
help to explain the phenomenology of short~$\boldsymbol{\gamma}$-ray bursts (sGRBs), and
nucleosynthesis processes powering kilonovae~\citep{Li:1998,Metzger:2016pju}.
To  understand these observations and, in particular, to understand the physics of matter
under extreme conditions, it is crucial to compare them to predictions from theoretical
modeling, which, due to the complexity of the underlying physical phenomena, is largely
numerical in nature.

Although a spinning BH surrounded by an accretion disk is the remnant of a BHNS merger,
this is not necessarily the case for an NSNS merger. Depending on the total mass of the system,
as well as the EOS of the NS companions, the outcome of an NSNS merger can be a stable NS or a
spinning BH, surrounded by an accretion disk in either case. Even when a BH is the remnant,
the path towards such an outcome is extremely varied and can be decisive for a number of
important issues, like the existence of a sGRB or the production of the heaviest elements
in the Universe via a kilonova~\citep{Metzger2014}. The current consensus for the event
  GW170817 is the formation of a transient
NS remnant sustaining itself for a brief period of time $\lesssim 1\,\rm s$ before collapsing to a BH 
(this was inferred from the existence of a sGRB, and the large amount of ejecta $\gtrsim 0.02
\,M_\odot$ estimated from the kilonova AT 2017gfo).
Assuming that this was the case, it is possible to place strong constraints
on the maximum mass of a cold spherical NS and its EOS 
\citep{Margalit2017,Shibata2017,Rezzolla_2018,Ruiz:2017due,Shibata2019}. 
These constraints could also provide an explanation for the unidentified $2.6\,M_\odot$
compact object in GW190814 as a rotating or even a nonrotating NS \citep{Most2020,Tsokaros2020}.
From a different point of view, the absence of a prompt collapse scenario and the large ejecta
mass also puts constraints on NS radii or, equivalently, their tidal
deformability~\citep{Bauswein:2017vtn,Radice:2017lry}. These constraints on the NS radius coming
directly from the postmerger object were further refined by complementary analyses of the GW
inspiral signal, which can be used to estimate the tidal deformability of the inspiraling NSs
\citep{TheLIGOScientific:2017qsa,Raithel2018,De2018}.

%Since only elements up to iron can form in thermodynamic equilibrium,
\cite{Lattimer74} and \cite{Symbalisty82} suggested that unstable neutron-rich nuclei
can be built in the mergers of BHNS or NSNS systems through rapid neutron bombardment,
the r-process. Apart from the 
dynamical ejecta that emerge within milliseconds after merger, the ejecta 
that emerge much later are very important in the determination of whether or not heavier
elements  through the r-process are being produced. \cite{Li:1998} argued that 
the low mass and high velocity of these ejecta  will make them transparent to their
own  radiation, resulting in emission whose peak will last around one day.
\cite{Metzger:2010} calculated the luminosity of the radioactively-powered transients in NS 
mergers and found these transients to be approximately 1000 times brighter than typical novae,
therefore calling them  ``kilonovae''. \cite{Metzger2014} argued that the lifetime of the merger
remnant is directly imprinted
in their early ``blue'' emission (from high electron fraction, lanthanide-poor ejecta) or late ``red''
emission (from low electron fraction, lanthanide-rich ejecta), both of which have been seen in
event GW170817.  The blue emission suggested ejecta composed of light r-process elements, while 
the red emission is consistent with heavier ones (lanthanide or actinides). The overall conclusion
is the the kilonova AT 2017gfo was a major source of r-process elements \citep{Kasen2017,Cote2017}. 

Another important characteristic associated with event GW170817 was the observation of an sGRB
-- event GRB 170817A~\citep{2017GCN.21517....1K,Monitor:2017mdv}. This GRB was unusually weak, and
various models have been proposed to explain this, including a choked-jet cocoon or a successful-jet cocoon 
\citep{Hallinan2017,Kasliwal2017,Mooley2018}.
Recently, \cite{Mooley2018} using radio observations from very long-baseline interferometry were
able to break the degeneracy between the choked and successful-jet cocoon models and concluded that
the early-time radio emission was powered by a wide-angle outflow (a cocoon), while the late-time
emission was most probably dominated by an energetic and narrowly collimated jet with an opening 
angle of less than five degrees, and observed from a viewing angle of about 20 degrees. This 
solidified theoretical predictions that NSNS, or at least a stellar binary where at least one
of the companions is a NS, can be the progenitors of the  central engine that 
power sGRBs \citep{p86,eetal89,npp92}.

Although GRB 170817A provided the long-sought observational evidence linking sGRBs with
NSNS mergers, it did not reveal the nature of the central engine behind the launching of a
relativistic jet. In particular, is a BH horizon necessary for the existence of a jet or is
it just sufficient~\citep{prs15,Ruiz:2018wah,Ruiz:2016rai,Ruiz:2017inq,Ruiz:2019ezy}? If necessary,
then a  stable NS remnant cannot be the generator of such jets. %, except in the case where
%the NS is close to the turning point line where a collapse to a BH is imminent.
If not,
%i.e NSs can launch themselves jets,
is the jet from a stable NS qualitatively the same as the one launched from a spinning BH immersed
in a gaseous disk?   In particular, can one  describe it as a \cite{BZeffect} (BZ) jet? 
Notice that according to~\cite{Komissarov:2002dj,Komissarov:2004ms,2005MNRAS.359..801K}
and~\cite{Ruiz:2012te},
the driving mechanism behind a BZ jet is not the horizon but the ergoregion.
Thus, while it may be that typical NSs~cannot launch a BZ jet, NSs that contain
ergoregions --ergostars-- might be able to~\citep{Ruiz:2020zaz}.

Since the pioneering general relativity (GR) simulations of NSNS mergers by~\cite{ShiU00} and BHNS 
mergers by~\cite{BSS},~\cite{Shibata:2006ks} and~\cite{FBST,FBSTR}, a number of groups have produced
a large body of work that captures the main characteristics of such events~(see reviews by
\cite{st11,Baiotti2016} and \cite{Foucart:2020ats}).
Below we will present a brief review of some of the important progress in the field, paying
special attention to pure hydrodynamical versus magnetohydrodynamical simulations.
Details regarding the  techniques used (either in evolution or in the initial data) will be
omitted. We refer the reader to e.g.~\cite{Alcubierre08a,BSBook,shibatabook} for such~details.
We also do not  treat white dwarf-neutron star (WDNS) mergers, which, though important for GW
detections by LISA, are not likely sources of sGRBs or kilonova. We refer readers interested
to the GR simulations of~\cite{Paschalidis:2011ez} and references therein.

We adopt geometrized units with $c=G=1$ unless otherwise indicated.

%%%%%%%%%%%%%%%%%%%%%
%%% BBHNS mergers %%%
%%%%%%%%%%%%%%%%%%%%%
\section{\bf Black hole-neutron star mergers: Remnants and incipient jets}
\label{sec:BHNS}
Motivated by the significance of BHNS binaries as copious sources of GW and EM radiation,
many numerical studies have been performed over the past years. Before the pioneering BHBH
simulations~\citep{FP1,RIT1,God1}, most dynamical simulations of BHNS binaries were
treated in Newtonian gravity, modeling the BH as a point mass~\citep{Lee00,RSW,Rosswog,Koba,
  RKLRasio}. Although these studies gave first insights on the basic dynamics of BHNSs, 
full  GR simulations  are required to properly model the late inspiral,
NS disruption, tidal tails, merger remnant, disk mass, fraction of unbound material
ejected, sGRB engine, and most significantly the GWs emitted during merger. In the following section,
we only review full GR studies of these binaries.

%
%%%%%%%%%%%%%%%%%%%%%%%%
%%% Hydro evolutions %%%
%%%%%%%%%%%%%%%%%%%%%%%%
\vspace{0.5cm}
\subsection{\bf Nonmagnetized evolutions}
\label{sec:hydro_BHNS}
Most of the close BHNS binary orbits are likely quasi-circular, since gravitational radiation reduces
the orbital eccentricity of the binary as it evolves toward smaller orbits
\citep{PhysRev.136.B1224}. However, a small fraction may form in dense stellar regions, such
as globular cluster or galactic nuclei, through dynamical capture, and they may merge with high eccentricities
\citep{Kocsis:2011jy,2010ApJ...720..953L,Samsing:2013kua}.

Motivated by the above, different groups have generated quasi-equilibrium initial data for
BHNSs on quasi-circular orbits~\citep{BSS,TBFS05,Shibata:2006ks,Shibata:2006bs,
  Grandclement:2006ht,TBFS07a,Foucart:2008qt}. Some of the earliest full GR simulations of
these configurations were performed by~\cite{Shibata:2006ks,Shibata:2006bs}, followed by
\cite{Etienne:2007jg} and~\cite{dfkpst08}. In all of these studies the binary was formed by a
nonspinning BH with a NS companion  modeled as a $\Gamma=2$ polytrope. These simulations
showed that the fate of BHNS remnants can be classified in two basic categories:
1) the NS is tidally disrupted before reaching the innermost stable circular orbit (ISCO),
inducing a long tidal tail of matter that eventually wraps around the BH and forms a
significant accretion disk (typically with a mass $\gtrsim 8\%$ of the NS rest-mass);
2) the NS plunges into the BH, leaving a BH surrounded by a negligibly small accretion disk
(typically with a mass $\lesssim 2\%$~of the NS rest-mass).

Using a Smoothed Particle Hydrodynamics (SPH) code and an approximate ``conformal'' GR metric,
\cite{RKLRasio} showed that the mass of the accretion disk remnant strongly depends on the
magnitude and direction of the BH spin. In particular, it was found that only systems with a
highly spinning BH, and slightly misaligned to the total angular of the system, yield significant
accretion disk remnants. These results were later confirmed by full GR studies \citep{Etienne:2008re,
  Foucart:2010eq,fdksst11,kost11} showing that for sufficiently high BH spins, mass ratios $q=M_{\rm BH}/
M_{\rm NS}\lesssim 3$, and/or lower NS compactions $\mathcal{C}={\mathcal M}_{\rm NS}/R_{\rm NS}~\lesssim 0.18$,
a substantial disk can form following merger. Here $M_{\rm BH}$ is the~\cite{1970PhRvL..25.1596C} BH
mass at infinite  separation and  $M_{\rm NS}$ the NS rest mass, while $\mathcal{M}_{\rm NS}$ and
$R_{\rm NS}$ are the gravitational (Arnowitt-Deser-Misner (ADM)) mass and the circumferential radius of the star in isolation,
respectively.

Using the above numerical simulation results, \cite{Foucart:2012nc} constructed a simple
fitting formula to predict the amount of matter remaining outside the BH horizon about
$10\,{\rm ms}$  following merger:
\begin{equation}
\frac{M_{\rm disk}}{M_{\rm NS}} \approx 0.415\,q^{1/3}\,\left(1-2\,\mathcal{C}\right) 
- 0.148\,\frac{R_{\rm ISCO}}{R_{\rm NS}}\,.
\label{eq:mass_esti}
\end{equation}
This expression is valid for mass ratios in the range~$q=3-7$, BH spins ~$a_{\rm BH}/M_{\rm BH}=
0-0.9$, and NSs with radii $R_{\rm NS}=11-16\,\rm km$, thereby encompassing the most likely
astrophysically
relevant parameter space. Here, $M_{\rm disk}$ and $R_{\rm ISCO}$ is the mass of the disk
remnant and the radius of the ISCO, respectively. Note that Eq.~\ref{eq:mass_esti} explicitly
shows that the mass of the disk remnant depends on the EOS and the BH spin,
which determine the mass and radius of the NS and the position of the ISCO, respectively.
It should be noticed that BHNSs with nearly-extremal BH spins have been considered by
\cite{lovelace08,Lovelace:2013vma}. These studies found that upon NS disruption, less than
half of the matter is promptly accreted by the BH, around $20\%$ becomes unbound and escapes,
and the remaining mass settles into a massive accretion disk.

Early population synthesis studies found that the
distribution of mass ratios in BHNSs depends on the metallicity and peaks at $q=7$
\citep{Belczynski:2007xg,Belczynski10}, but more recent works found that it is generally
less than $10$, peaking at $q\approx 5$~\citep{Giacobbo:2018etu,Abbott:2020khf}.
Using Eq.~\ref{eq:mass_esti}, one finds that, for a binary with mass ratio $q=5$
in which the  NS companion has radius~$13.3\,\rm km$ and rest-mass $M_{\rm NS}=1.44\,M_\odot$
(compatible with NICER observations;~\cite{Miller_2019,Riley:2019}) a BH spin of~$a_{\rm BH}/
M_{\rm BH} \gtrsim 0.65$ is required  to form an accretion disk with~$\gtrsim 10\%$ of the NS rest mass.  
The power available for EM emission is usually taken to be proportional to the accretion
rate. Under this assumption, it is expected that the luminosity of the disk remnant
is $L_\mathrm{EM} = \epsilon\,\dot{M}_\mathrm{disk}$, where $\epsilon$ is the efficiency
for converting accretion power to EM luminosity and $\dot{M}_\mathrm{disk}\sim
M_{\rm disk}/t_{\rm acc}$ is the rest-mass accretion rate, where $t_{\rm acc}$ is the disk
lifetime. Assuming a $1\%$ efficiency and a disk lifetime of $\sim 0.2\,\rm s$, the luminosity
is $L_\mathrm{EM} \sim 10^{51}\rm erg/s$,  consistent
with typical EM luminosities of sGRBs. This value is also consistent with the ``universal'' merger
scenario for generating EM emission from merger and collapse  BH + disk remnants~\citep{shapiro17}.
These results allow us to conclude that the merger of NSs orbiting highly spinning BHs can be the
progenitors of the engines that power sGRBs.
However, the LV scientific collaboration has reported the observation of
BHBHs having high mass and/or low spins (see e.g.~Table VI in~\cite{Abbott:2020niy}).
If this trend continues for LV-like BHNSs, then it is expected that LV-like BHNS remnants
would have negligible accretion disks, which might disfavor their role as progenitors
of sGRBs and/or observable kilonovae.

The previous numerical studies assumed that the NS companion is irrotational. Recently,~\cite{East:2015yea}
and~\cite{Ruiz:2020elr} showed that the NS spin has a strong impact on the disk remnant and the dynamical
ejecta. As the prograde NS spin increases, the effective ISCO decreases
\citep{Barausse:2009xi}. In addition, as the magnitude of the NS spin increases, the star becomes
less bound and the tidal separation radius $r_{\rm tid}$ (separation at which tidal disruption begins)
increases, also resulting in more pronounced disruption effects. This effect can be easily understood
by estimating~$r_{\rm tid}$ by equating the inward gravitational force exerted by the NS on its fluid
elements with the BH's outward tidal and the outgoing centrifugal forces to obtain
\begin{equation}
  r_{\rm tid}/M_{\rm BH}\simeq q^{-2/3}\,\mathcal{C}^{-1}\,
  \left[1-\Omega^2\,M_{\rm NS}^2\,\mathcal{C}^{-3}\right]^{-1/3}\,,
 \label{eq:tidal}
\end{equation}
\citep{Ruiz:2020elr} where $\Omega=a_{\rm NS}M_{\rm NS}/I$. Here $a_{\rm NS}$ is the NS spin parameter
and $I$ its moment of inertia. This simple Newtonian expression
shows that the larger the mass ratio and/or the compaction of the NS, the closer the tidal
separation to the ISCO. The NS then experiences tidal disruption effects only during a short
time before the bulk of the NS plunges onto the BH. In contrast, the larger the magnitude of the NS
spin, the farther away $r_{\rm tid}$ is from  the ISCO. In this case, the star can be tidally disrupted
before being swallowed by the BH which increases the time for disruption and with it the amount of matter
that spreads out to form the disk or escapes to infinity.

Recently, \cite{Barnes:2013wka} showed that the opacities in r-process ejecta are likely dominated
by lanthanides, which induce peak bolometric luminosities for kilonovae of
\begin{equation}
  L_{\rm knova}\approx 10^{41}\left(\frac{M_{\rm eje}}
  {10^{-2}M_{\odot}}\right)^{1/2}\,\left(\frac{{\rm v}_{\rm eje}}{0.3c}\right)^{1/2}\,
  \rm erg/s\,,
 \label{L_peak_knove}
\end{equation} 
\citep{East:2015yea}~and rise times of
\begin{equation}
  t_{\rm peak}\approx0.25\,\left(\frac{M_{\rm eje}}{10^{-2} M_{\odot}}\right)^{1/2}\,
  \left(\frac{{\rm v}_{\rm eje}}{0.3c}\right)^{-1/2}\, \rm days\,,
\label{t_peak_knove}
\end{equation}
\citep{East:2015yea} where ${\rm v}_{\rm eje}$ and $M_{\rm eje}$ are the mass-averaged velocity and rest-mass of the ejecta.
The characteristic speed of the ejecta is ${\rm v}_{\rm eje}/c\lesssim 0.2-0.3$ with a rest-mass of
$\lesssim 10^{-3}\,M_\odot$~(see e.g.~\cite{East:2015yea,Ruiz:2020elr,Foucart:2019bxj,Hayashi:2020zmn}).
Therefore, the bolometric luminosity of kilonova signals is  $L_{\rm knova}\lesssim 10^{41}
\,\rm erg/s$ with rise times of $\lesssim 7\,\rm h$. These luminosities correspond to an R band magnitude
of $\sim 24$ mag at $200$~Mpc inside the aLIGO volume~\citep{Aasi:2013wya}, and above the LSST
survey sensitivity of $24.5$ mag~\citep{Barnes:2013wka,East:2015yea}. Hence some of these
signals may be detectable by the LSST survey~\citep{Ruiz:2020elr}. 

%
%%%%%%%%%%%%%%%%%%%%%%%%%%%%%
%%% Magnetized evolutions %%%
%%%%%%%%%%%%%%%%%%%%%%%%%%%%%
\hspace{0.5cm}
\subsection{\bf Magnetized evolutions}
\label{sec:magnetized_BHNS}
The previous numerical studies showed that BHNS mergers can create the right conditions to power
sGRBs (i.e. a~spinning BH + disk). However, they do not account for either magnetic fields or
neutrino pair annihilation processes, the most popular components invoked in most sGRB models
to drive jets~(see~e.g.~\cite{BZeffect,Vlahakis2003,SGRB_review,Aloy:2004nh}). 
As the lifetime of the neutrino pair annihilation process  might be too small to explain
typical sGRBs~\citep{Kyutoku:2017voj}, we henceforth focus only on the magnetic 
process. However, it is worth noting that BH + disk remnants powering sGRBs may be dominated
initially by neutrino pair annihilation processes followed by the BZ
mechanism~\citep{Dirirsa:2017pgm}, leading to a transition from a thermally-dominated fireball
to a Poynting EM-dominated flow, as is inferred for some GRBs, such as GRB
160625B~\citep{2018NatAs...2...69Z}.

Ideal GR magnetohydrodynamics (GRMHD) studies of magnetized BHNS mergers in which the NS
is initially endowed with an {\it interior-only} poloidal magnetic field generated by the
vector potential
\begin{eqnarray}
  A_i = \left( -\frac{y-y_c}{\varpi_{\rm c}^2}\delta^x{}_i 
+ \frac{x-x_{\rm c}}{\varpi^2_{\rm c}}\delta^y{}_i\right)\,A_\varphi\,, \hspace{0.5cm} 
  A_\varphi = A_b\,\varpi^2_{\rm c}\,\max(P- P_{\rm cut},0)^{n_b}\,, 
\label{ini:Aphi_int}
\end{eqnarray}
were carried out by~\cite{Chawla:2010sw},~\cite{Etienne:2011ea} and~\cite{Kiuchi:2015qua},
varying the mass ratio, the BH spin, and the strength of the magnetic field.
Here the orbital plane is at $z=0$, $(x_{\rm c},y_{\rm c},0)$ is the coordinate location of
the center of mass of the NS,
$\varpi^2_{\rm c}=(x-x_{\rm c})^2+(y-y_{\rm c})^2$, and $A_b$, $n_p$ and $P_{\rm cut}$ are free
parameters. The cutoff pressure parameter $P_{\rm cut}$ confines the magnetic field
inside the NS within $P>P_{\rm cut}$. The parameter $n_b$ determines the degree
of central condensation of the magnetic field.
%
%%%%%%%%%%%%%%%%%
%%%  Etienne  %%%
%%%%%%%%%%%%%%%%%
\begin{figure}[h!]
\begin{center}
  \includegraphics[width=5.9cm]{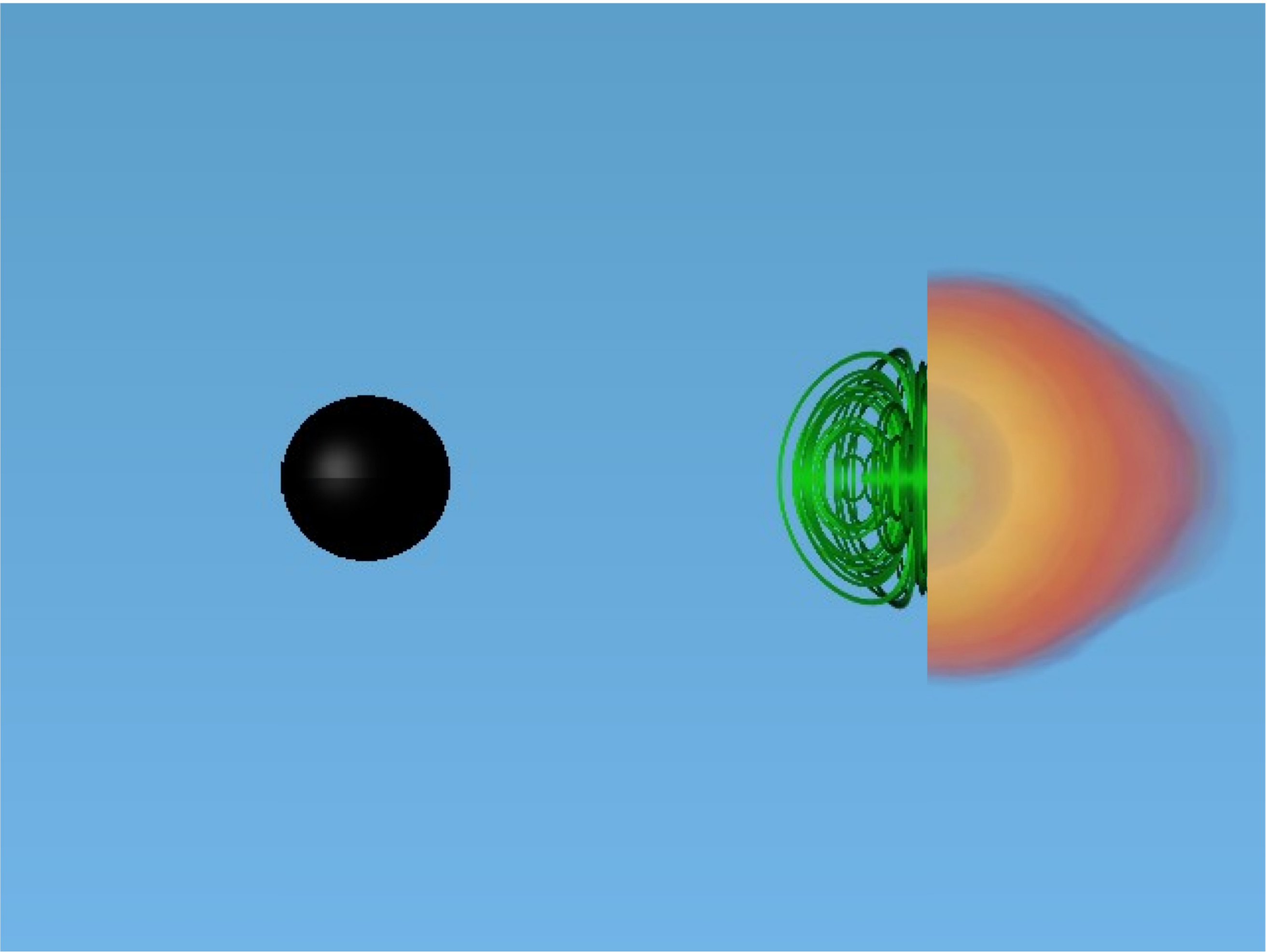}
  \includegraphics[width=5.9cm]{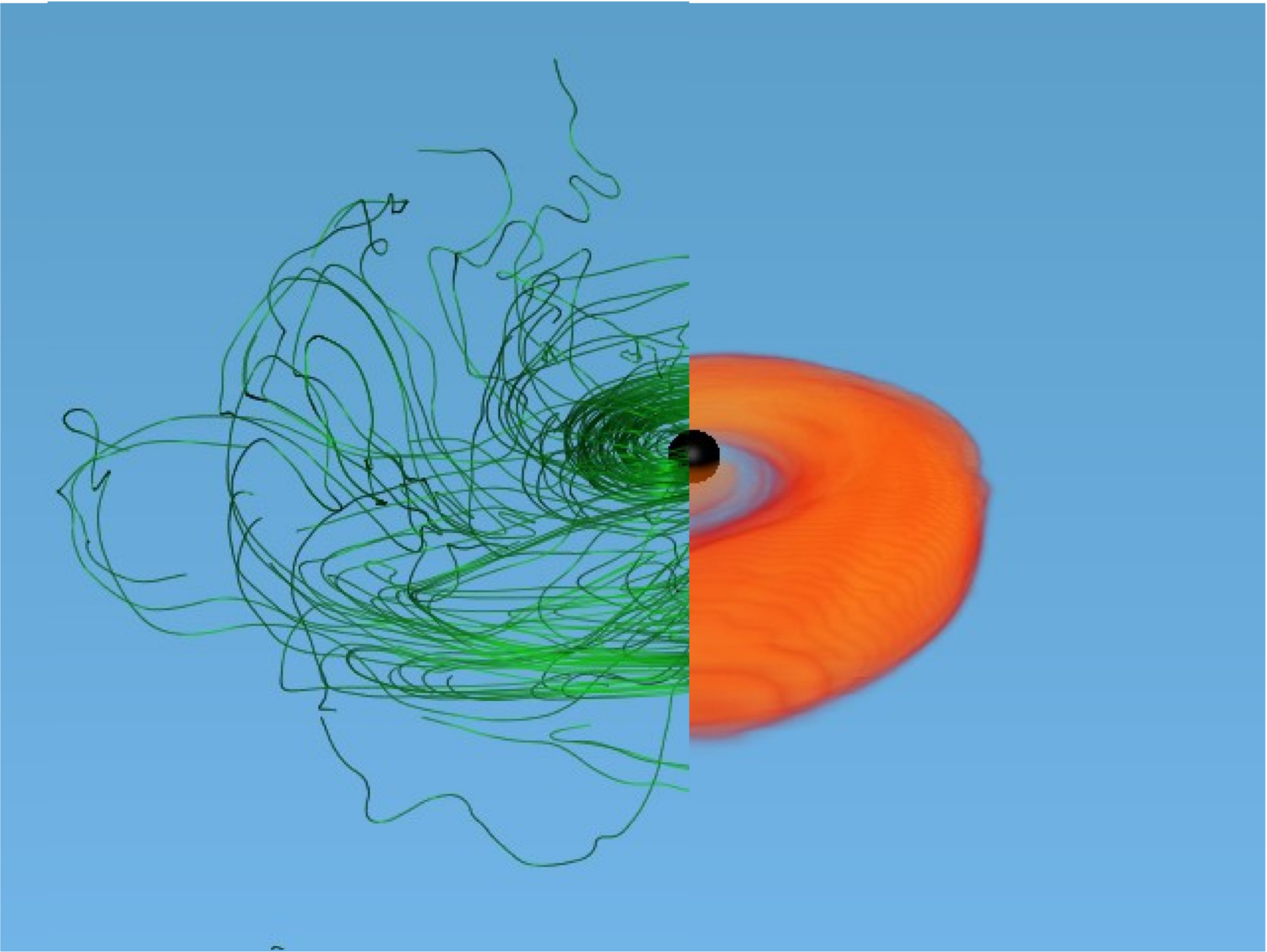}
  \includegraphics[width=5.9cm]{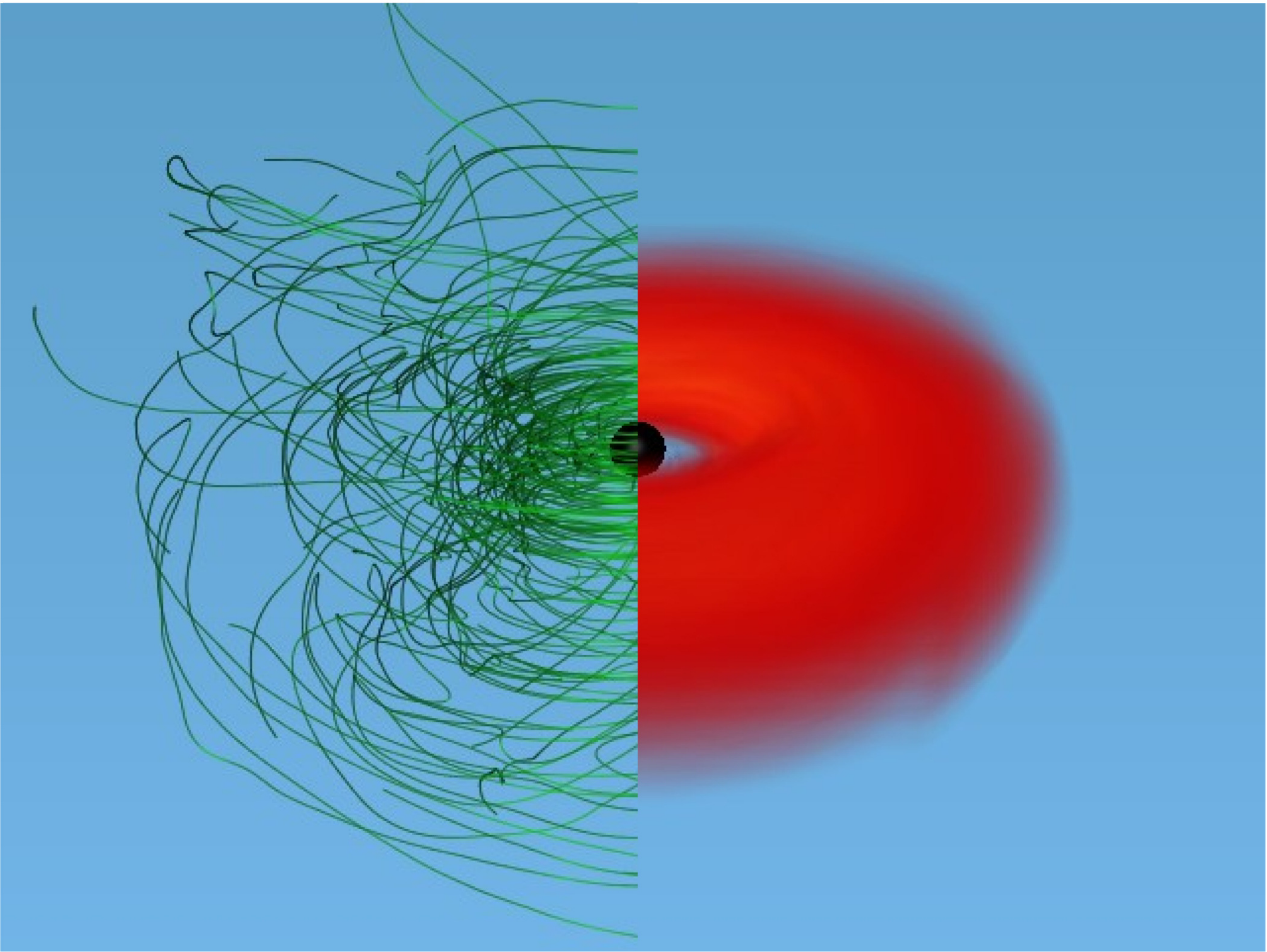}
\end{center}
\caption{The NS magnetic field lines (green) and rest-mass density $\rho_0$ (reddish) normalized to
  the initial NS maximum value $\rho_0=8.92\times 10^{14}\,(1.4M_\odot/M_{\rm NS})^2\rm{g\,/cm}^{3}$,
  at selected times for a BHNS  with mass ratio~$q=3$.  The initial BH spin is $a_{\rm BH}/M_{\rm BH}
  =0.75$ and the NS is an irrotational $\Gamma=2$ polytrope. Here the BH apparent horizon is shown as
  a black sphere. Following merger, the field lines are wound into an almost purely toroidal configuration
  [Adapted from~\cite{Etienne:2012te}].}
\label{fig:Etienne2012}
\end{figure}

\begin{figure}[h!]
\begin{center}
  \includegraphics[width=16.0cm]{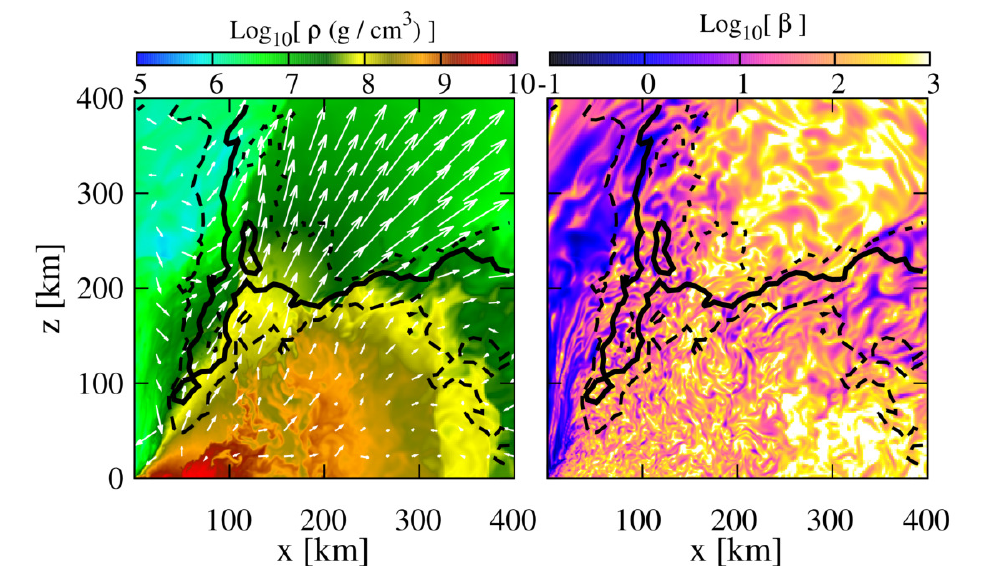}
\end{center}
\caption{NS rest-mass density 
  with fluid velocity arrows (left) and the gas-to-magnetic-pressure
  ratio (right) of a $q=4$ BHNS remnant after $\sim 50\,\rm ms$ following merger.
  A thermally-driven wind (but no collimated) outflow is observed [From~\cite{Kiuchi:2015qua}].}
\label{fig:Kiuchi_2015}
\end{figure}

These numerical simulations showed that following merger, tidal tails of matter wrap around the BH,
forming the accretion disk and dragging the frozen-in magnetic field into an almost purely toroidal
configuration (see~Fig.~\ref{fig:Etienne2012}). These simulations did not find any evidence of jet
launching following the BH + disk formation. Nevertheless,~\cite{Kiuchi:2015qua} reported that
in their high-resolution simulations, in
which the mass ratio is $q=4$, the BH has a spin $a_{\rm BH}/M_{\rm BH}=0.75$, and the NS is modeled
by the APR EOS~\citep{Akmal:1998cf}, a thermally-driven wind (but no collimated) outflow emerges
after $\sim 50\,\rm ms$ following merger~(see~Fig.~\ref{fig:Kiuchi_2015}). %This outflow induces the
%formation of a large-scale poloidal magnetic field component.

The lack of magnetically-driven jets in these simulations has been attributed to the fact that the magnetic
field in the disk remnant is almost purely {\it toroidal}.~\cite{GRMHD_Jets_Req_Strong_Pol_fields} showed
that BH + disk systems can launch and support magnetically-driven  jets  only if a net {\it poloidal} magnetic
flux is accreted onto the BH. Motivated by this conclusion,~\cite{Etienne:2012te} endowed the disk remnant
from an unmagnetized BHNS simulation with a purely poloidal field and found that, indeed, under the
{\it right conditions},
a jet can be launched from BHNS remnants. However, identifying the initial configuration of the seed magnetic
field in the NS prior to tidal disruption that could lead to these conditions remained elusive for many years.

\cite{prs15} then demonstrated that a more realistic initial magnetic configuration for the NS companion 
--a dipolar magnetic field extending from the NS interior into the {\it exterior} (as in pulsars)-- could do the
trick. Such a field can be generated by the vector potential
\begin{equation}
  A_\phi= \frac{\pi\,\varpi^2\,I_0\,r_0^2}{(r_0^2+r^2)^{3/2}}
  \left[1+\frac{15\,
    r_0^2\,(r_0^2+\varpi^2)}{8\,(r_0^2+r^2)^2}\right]\,, 
\label{eq:Aphi}
\end{equation} 
\citep{Paschalidis:2013jsa}~which approximately corresponds to a vector potential generated by an interior current loop.
Here $r_0$ is the current loop radius, $I_0$ is the current, and~$r^2=\varpi^2+z^2$,
with $\varpi^2 =(x-x_{\rm c})^2 +(y-y_{\rm c})^2$.
%and $(x_{\rm NS},y_{\rm NS})$ is the position of the center of mass of the NS.
To reliably evolve  the  exterior magnetic field with an ideal GRMHD code %(in this the Illinois GRMHD
%code)
and simultaneously mimic the magnetic-pressure dominant environment that characterizes a pulsar-like
magnetosphere, a low and variable density atmosphere was installed initially in the exterior where
magnetic field stresses dominate over the fluid pressure.
%
%%%%%%%%%%%%%%%%%%%%%
%%%  Kiuchi_BHNS  %%%
%%%%%%%%%%%%%%%%%%%%%

\begin{figure}[h!]
  \begin{center}
    \includegraphics[width=5.9cm]{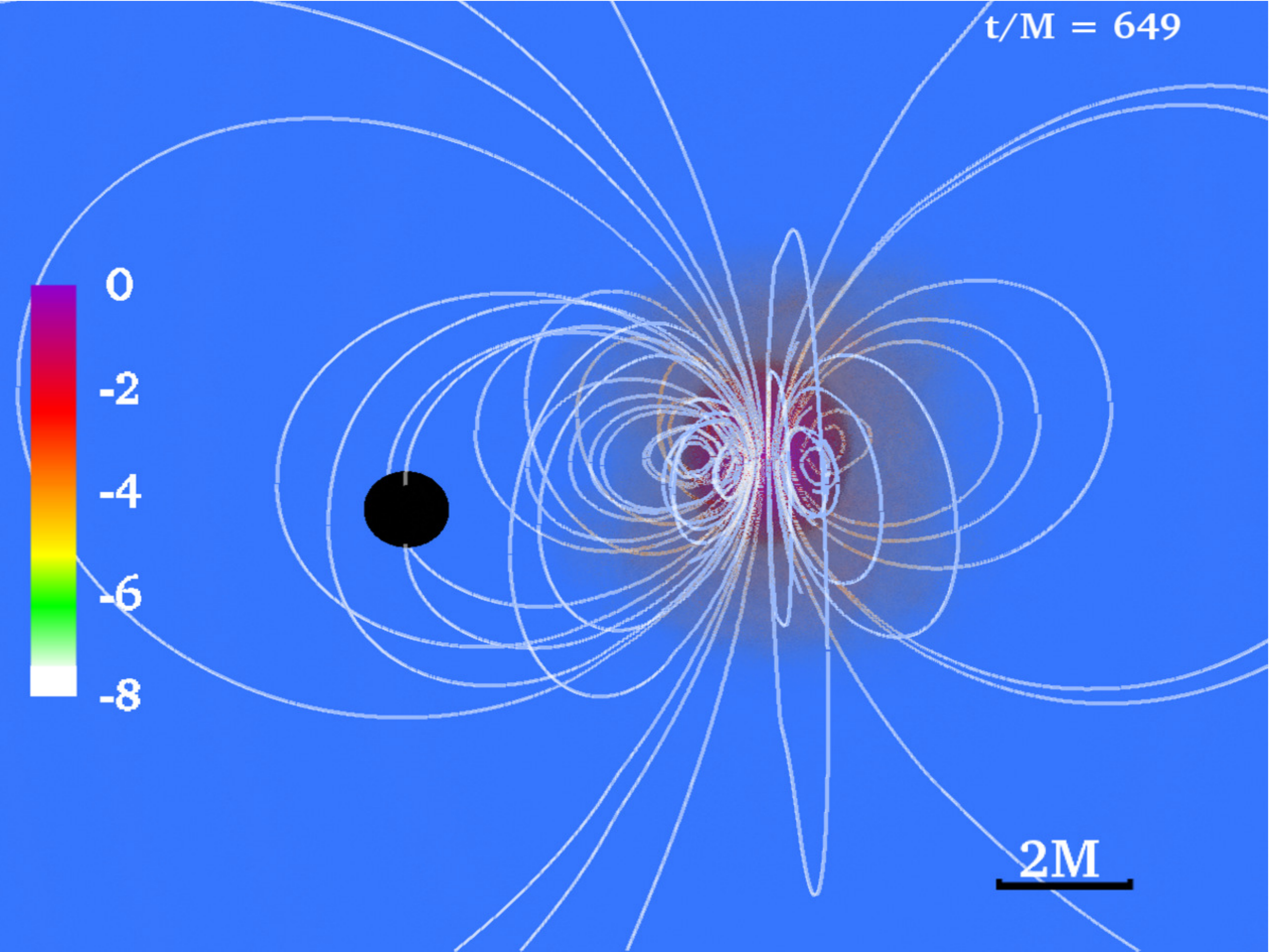}
    \includegraphics[width=5.9cm]{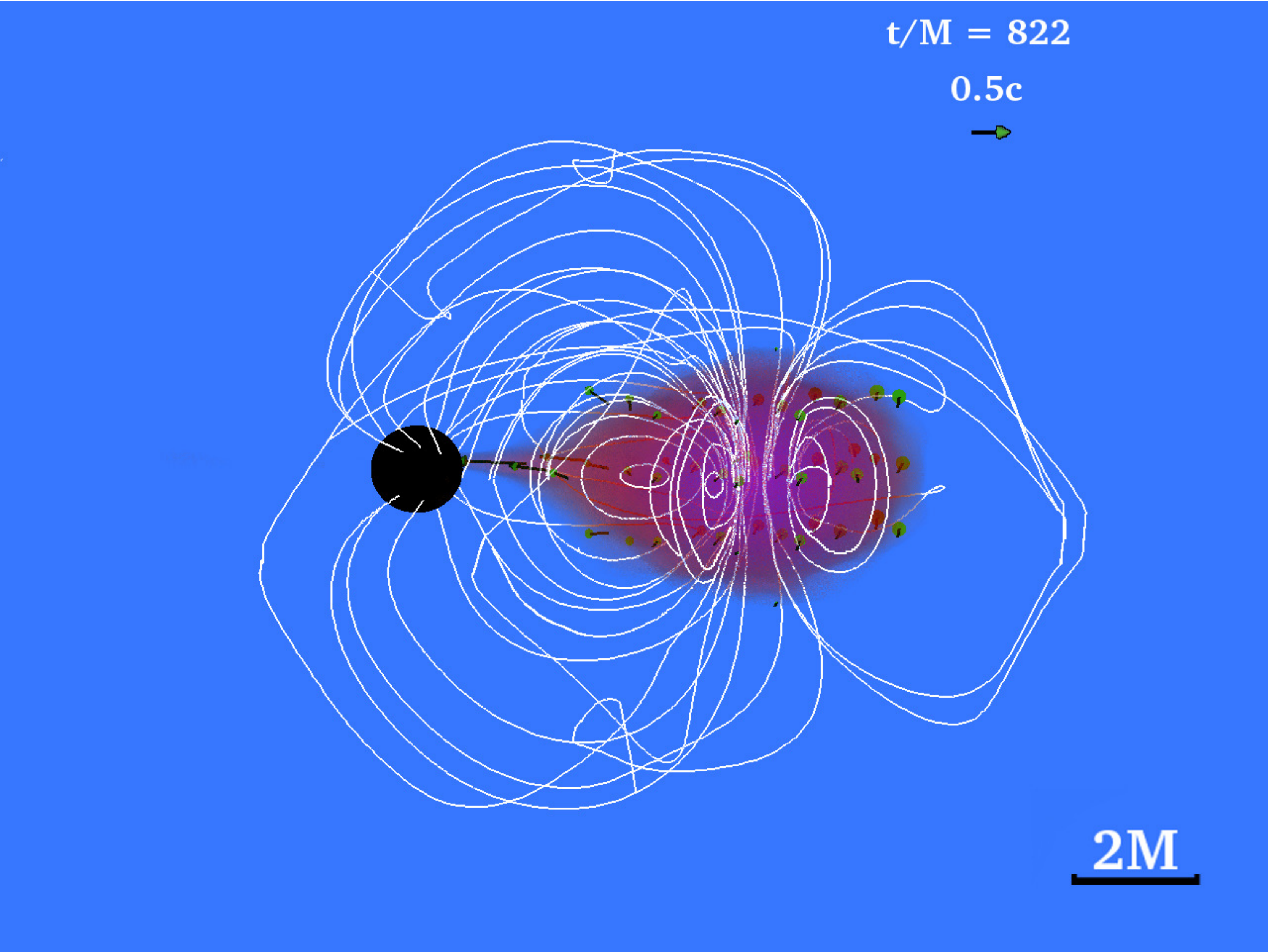}
    \includegraphics[width=5.9cm]{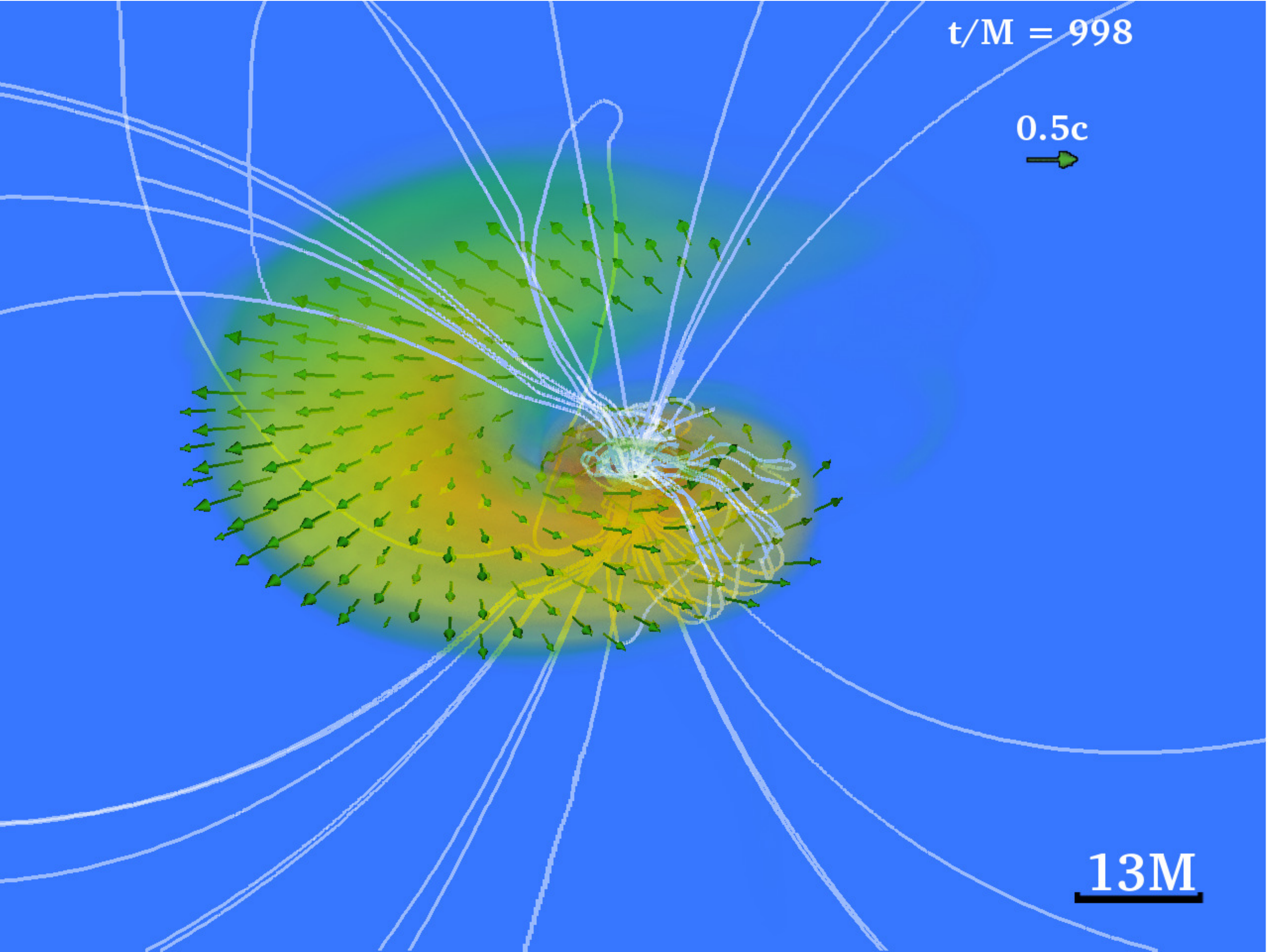}
    \includegraphics[width=5.9cm]{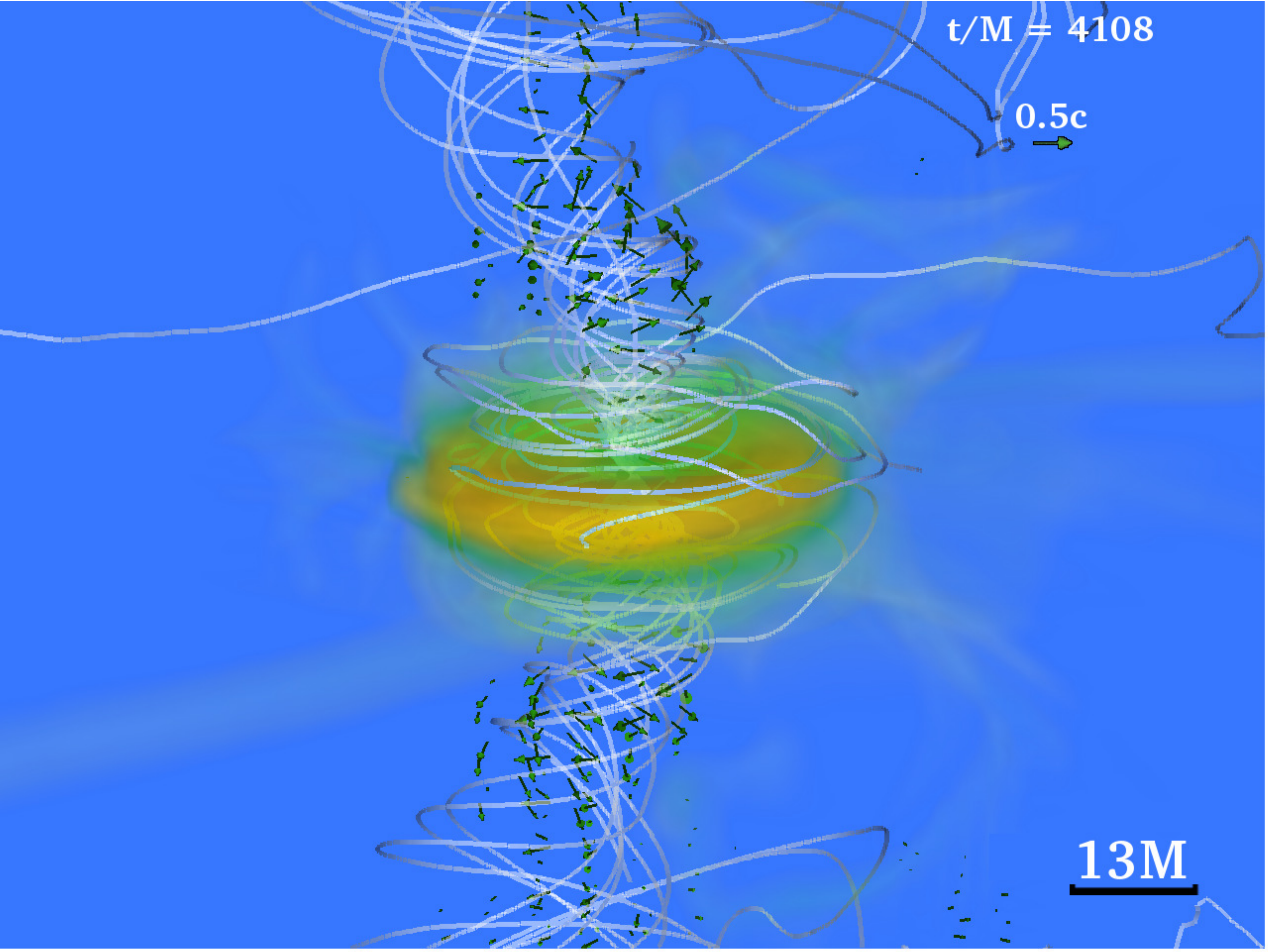}
    \includegraphics[width=5.9cm]{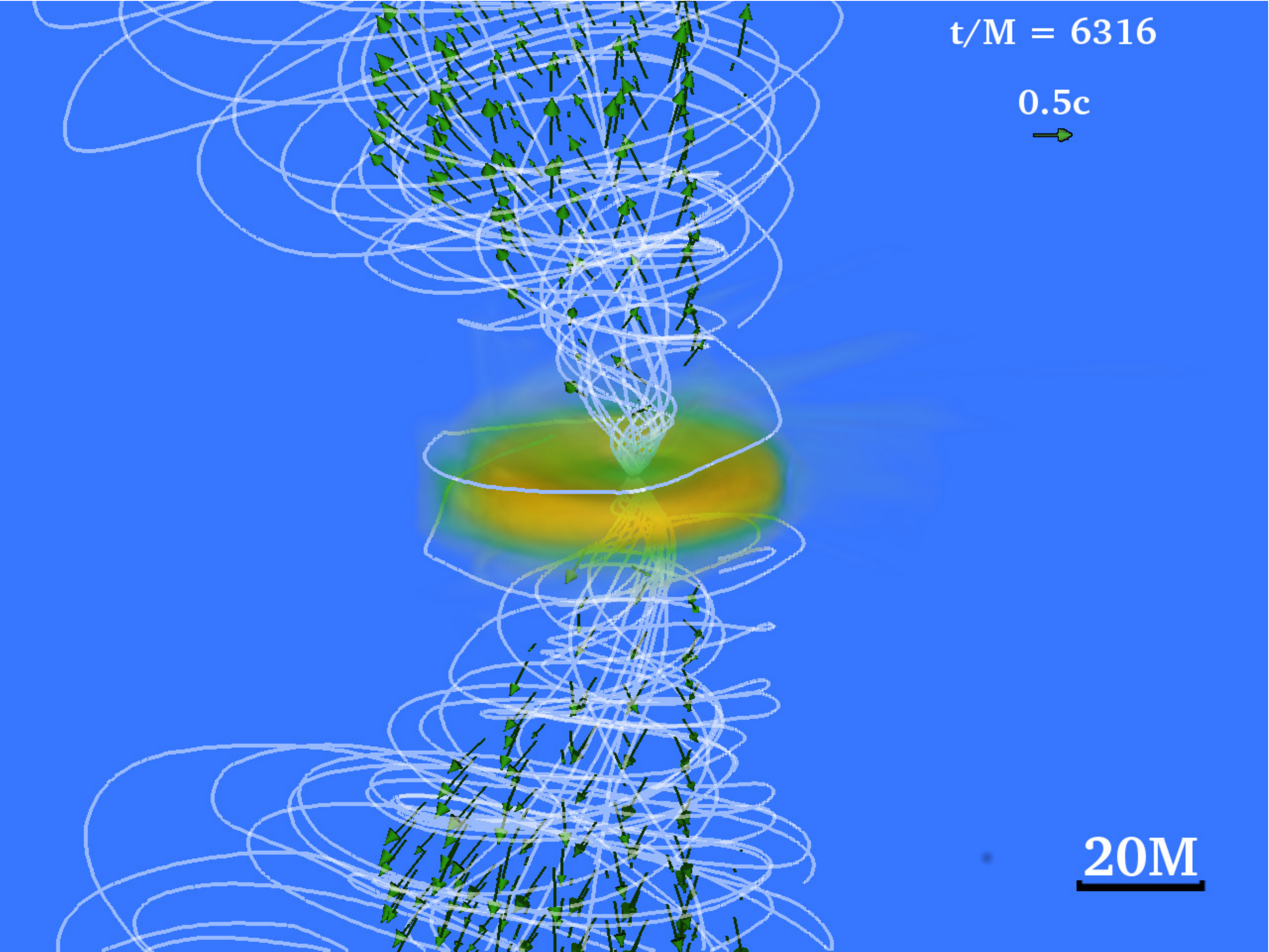}
    \includegraphics[width=5.9cm]{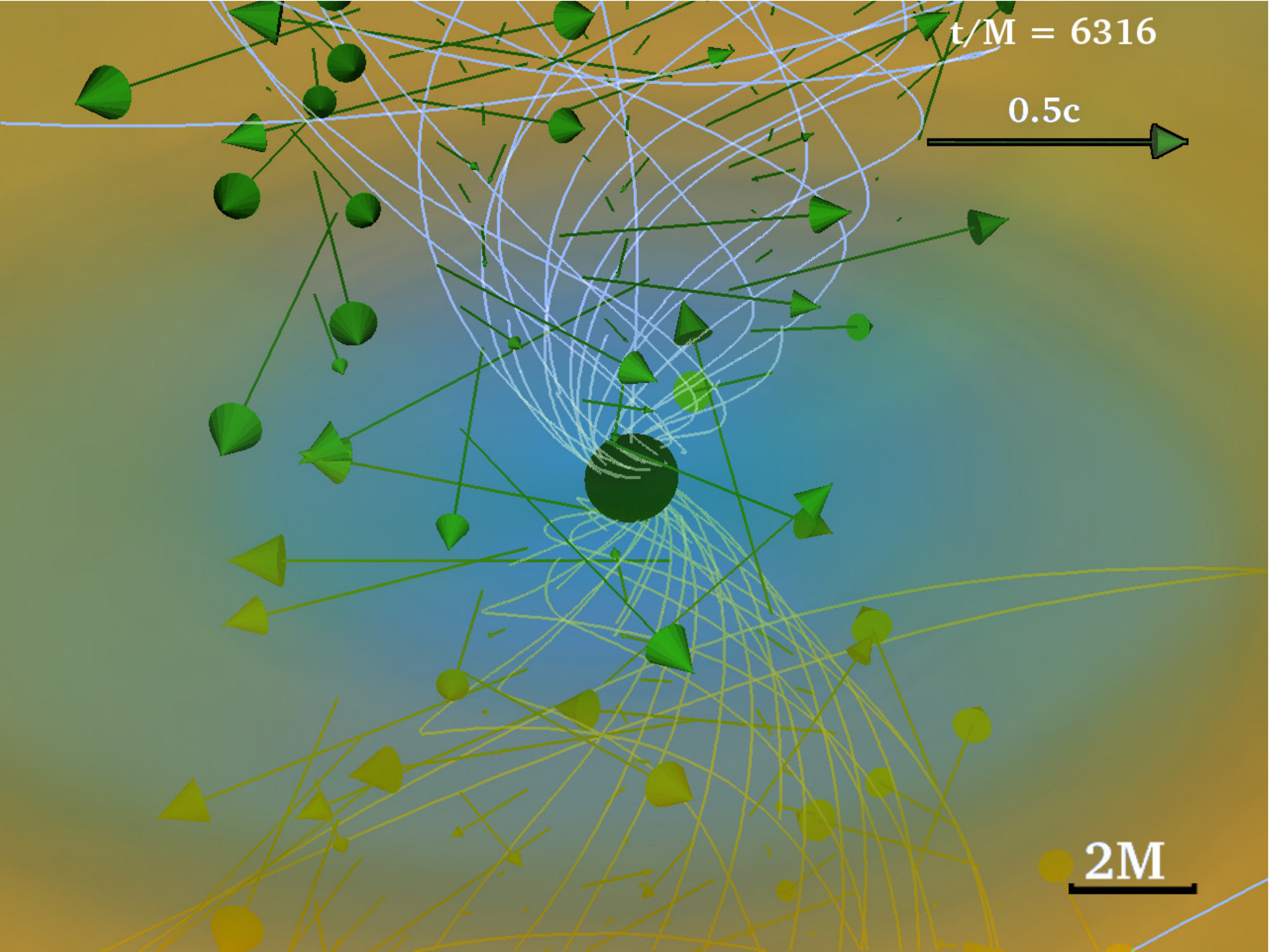}    
\end{center}
  \caption{NS rest-mass density $\rho_0$ normalized to its initial maximum value~$\rho_{0,\rm max}=8.92\times
    10^{14}(1.4M_\odot/M_{\rm NS})^2\,\rm g/cm^3$ (log scale) at selected times for a BHNS with mass ratio
    $q=3$. The initial BH spin is $a_{\rm BH}/M_{\rm BH}=0.75$ and the NS is an irrotational $\Gamma=2$
    polytrope. Arrows indicate fluid velocities and white lines
    the magnetic field lines. Bottom  panel shows  the system after an incipient jet is launched.
    Here ${\rm M}=2.5\times 10^{-2}(M_{\rm NS}/1.4M_\odot)\,\rm ms =7.58(M_{\rm NS}/1.4M_\odot)\,\rm km$
   [From~\cite{prs15}].}
  \label{fig:paschalidis_2014}
\end{figure}

The above technique was used by~\cite{prs15} and~\cite{Ruiz:2018wah,Ruiz:2020elr} to perform a series
of BHNS simulations varying the density of the ``artificial'' atmosphere, the binary mass-ratio, the
BH and NS spins, and the orientation of the seed magnetic field axis with respect to the orbital angular
momentum. It was found that independent of the atmosphere or the NS spin, a magnetically driven, incipient
jet is launched  once the regions above the BH poles become nearly force-free ($B^2/8\,\pi \rho_0\gg 1$)
for small tilt-angle magnetic fields and binary mass ratios that yield a significant disk remnant.
The jet is confined  by a collimated, tightly wound, helical magnetic funnel above the BH poles.
Following the onset of accretion, the magnetic field in the disk remains predominantly toroidal as in
the previous simulations. However, the external magnetic field maintains a strong poloidal component
that retains footpoints at the BH poles.
Magnetic instabilities (mainly magnetic winding and magnetorotational (MRI))  amplify the magnetic field
from $\sim 10^{13}(1.4M_\odot/M_{\rm NS})\,\rm G$ to $\sim 10^{15}(1.4M_\odot/M_{\rm NS})\,\rm G$
at the BH poles, and after $\Delta t\sim 90-150(M_{\rm NS}/1.4M_\odot)\,\rm ms$ following merger a
bonafide jet finally emerges~(see Fig.~\ref{fig:paschalidis_2014}). It is worth noting that the calculation
of~\cite{Ruiz:2020elr} showed that the larger the initial NS prograde spin, the larger the mass of the
accretion disk remnant. Similar behavior was observed for the amount of unbound ejecta. These results
suggest that moderately high-mass ratio BHNSs  ($q\lesssim 5$) that undergo merger, where the NS
companion has a non-negligible spin, may give rise to detectable kilonovae even if
magnetically-driven jets are not formed.

The Lorentz factor in the funnel~is~$\Gamma_{\rm L}\sim 1.2-1.3$,~and~hence the jet just above the BH poles
is only mildly relativistic. However, the maximum attainable Lorentz factor of a magnetically--powered, nearly
axisymmetric jet is comparable to the force-free parameter $B^2/8\pi\rho_0$~inside the funnel~\citep{Vlahakis2003}.
Near the end of the simulations the force-free parameter in the funnel reaches values $\gtrsim 100$. Thus,
it is expected that the jet will be accelerated to~$\Gamma_{\rm L}\gtrsim 100$ as required by most sGRB
models~\citep{Zou2009}.
%
%%%%%%%%%%%%%%%%%%%%%%%%%%
%%%  Paschalidis_2014  %%%
%%%%%%%%%%%%%%%%%%%%%%%%%%

The lifetime of the disk is $\Delta t\sim 500(M_{\rm NS}/1.4M_\odot)-700(M_{\rm NS}/1.4M_\odot)\,\rm ms$
and the outgoing EM Poynting luminosity is $L_{\rm EM}\sim 10^{51\pm 1}\rm erg/s$, and hence consistent with typical
sGRBs~\citep{Bhat:2016odd,Lien:2016zny,Svinkin:2016fho,Ajello:2019zki}. The luminosity is also consistent with
that generated  by  the BZ mechanism
 \begin{equation}
   L_{\rm BZ}\sim10^{51}\,\left(\frac{a_{\rm BH}}{M_{\rm BH}}\right)^2\,\left(\frac{M_{\rm BH}}{5.6M_\odot}\right)^2\,
   \left(\frac{B}{10^{15}\rm G}\right)^2 \rm erg\,/ s\,,
   \label{eq:LBZ}
 \end{equation}
\citep{Thorne86} as well as with the simple analytic  model that seems  to apply universally for typically  compact binaries
 mergers containing magnetized NSs that leave BH + disk remnants~\citep{shapiro17}. 

 The above results were obtained with a high initial magnetic field. \cite{prs15} argued %\sout{that is expected}
 that a smaller initial field will yield the same qualitative  outcome because the magnetic field amplification 
 following disruption is due largely  to magnetic winding and the MRI. Amplification proceeds  until
 appreciable differential rotational and internal energy of the plasma in the disk
 %, which is insensitive to the initial field,
 has been converted  to magnetic energy.
 This amplification yields~$B\sim~10^{15}G$ at the BH poles nearly independent of the
 initial NS magnetic field. Winding occurs on an Alfv\'en timescale, so  amplification may take longer the weaker  the
 initial field.

\begin{figure}[h!]
  \begin{center}
    \includegraphics[width=8.5cm]{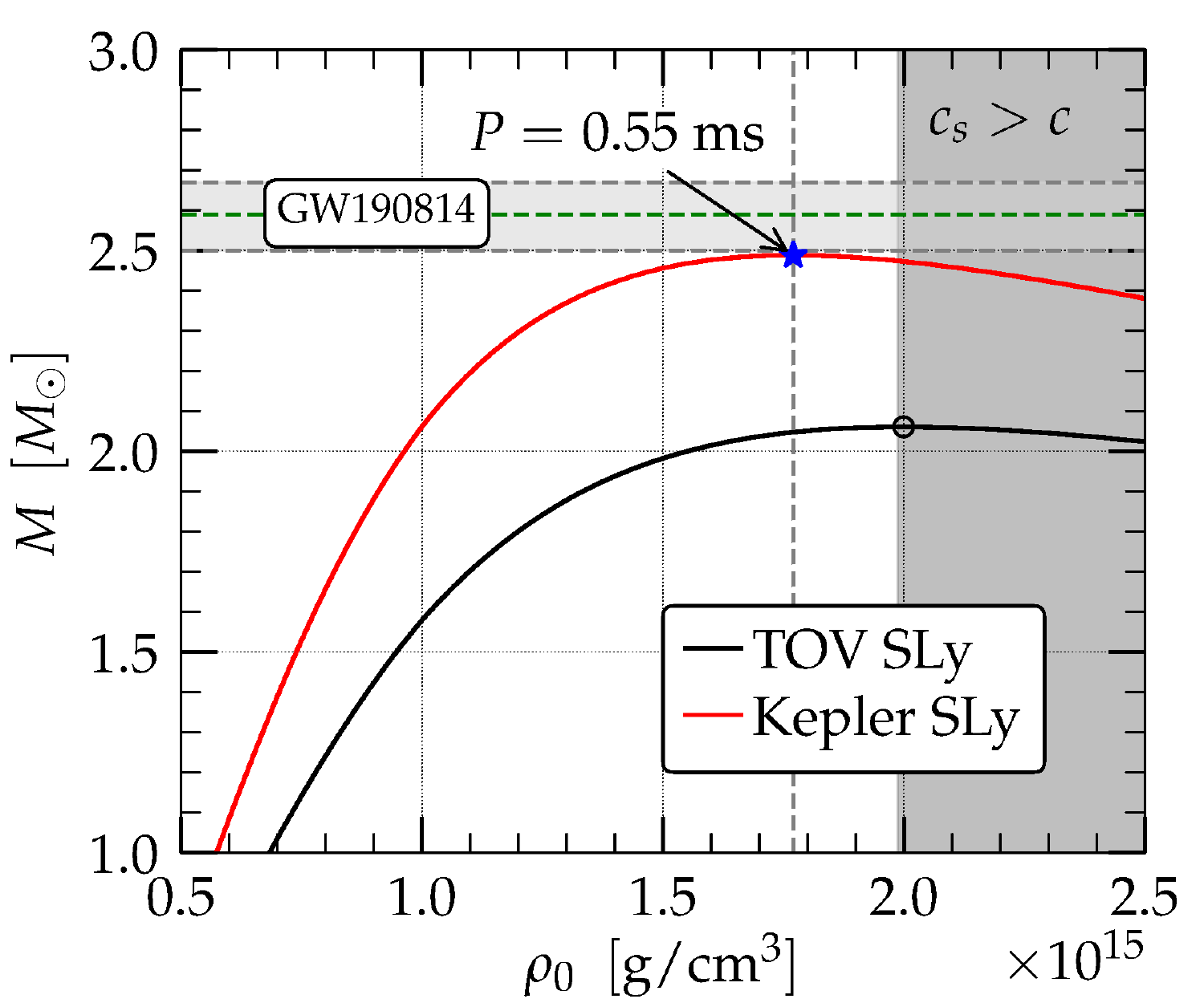}
    \includegraphics[width=8.5cm]{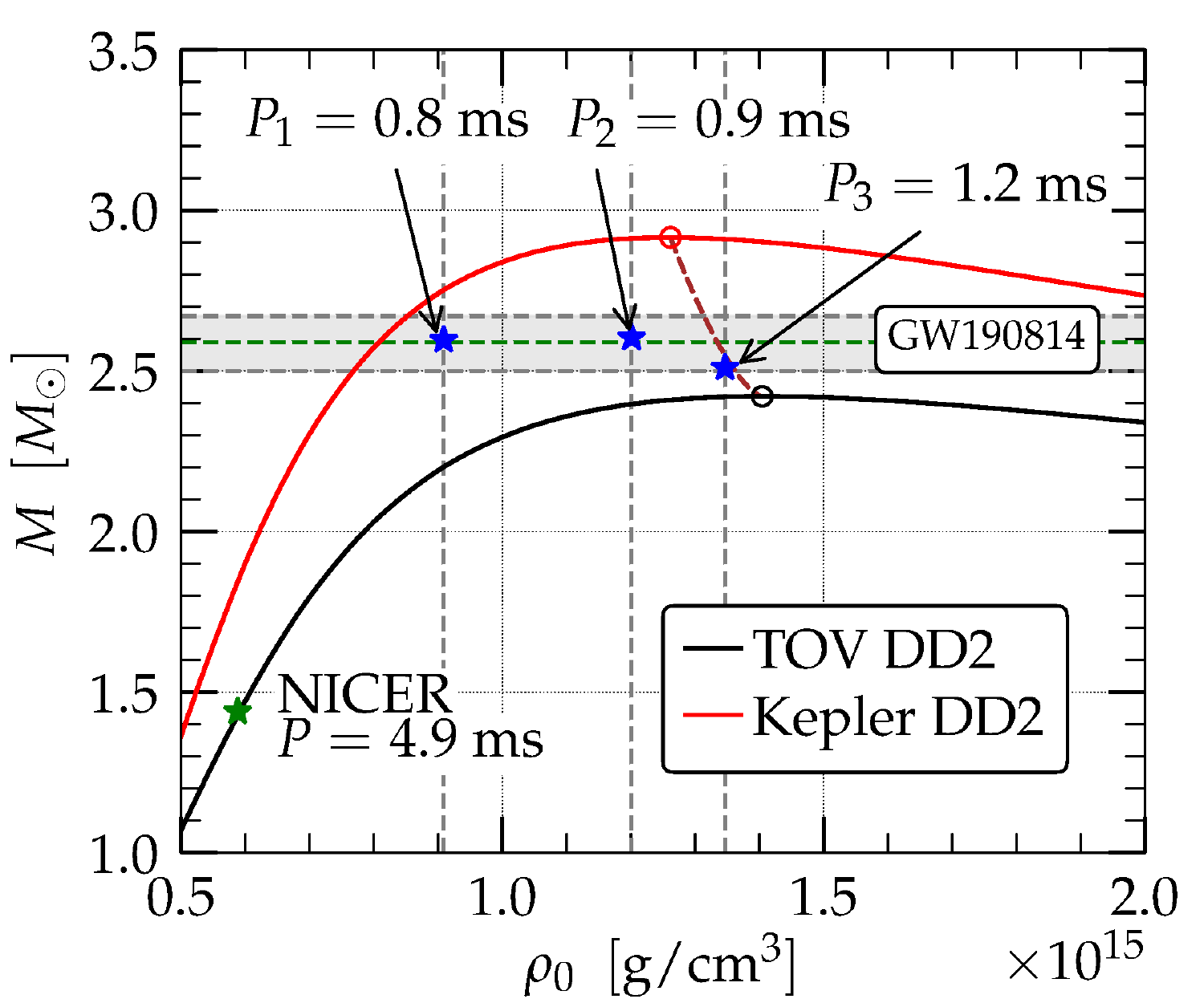}
  \end{center}
\caption{Two possibilities for the EOS of a NS companion  in GW190814. 
  The scenario on the left, which employs the SLy (soft) EOS, fails to provide a model
  for a uniformly rotating star, even at maximum uniform rotation. On the contrary, the
  scenario on the right that employs the DD2 (stiff) EOS succeeds and demonstrates 
  the possibility of a slowly rotating NS. The lower (black) curves represent
  spherical, nonrotating models, while the upper (red) curves represent uniformly rotation
  models spinning at the Keplerian (mass-shedding) limit~[From~\cite{Tsokaros2020}].}
\label{fig:Tsokaros_gw19}
\end{figure}

%%%%%%%%%%%%%%%%%%%
%%%  GW190814   %%%
%%%%%%%%%%%%%%%%%%%
\vspace{0.5cm}
\subsection{\bf GW190814: Spin \& EOS for a NS companion}
One of the most intriguing GW detections to date was event GW190814 \citep{Abbott:2020khf}, 
a binary coalescence whose primary component had mass $m_1=23.2_{-1.0}^{+1.1}\,M_\odot$ and
therefore is a BH, while the secondary had mass $m_2=2.59_{-0.09}^{+0.08}\,M_\odot$, placing it at
the boundary of the so-called ``mass gap'' and making its identification uncertain.
Further ambiguity was added by the absence of an EM counterpart. While the nature of this
compact object is not yet known, it was already suggested by~\cite{Abbott:2020khf}
that it can be a rapidly rotating NS, whose dimensionless spin was estimated to be $0.49
\lesssim a_{\rm NS}/M_{\rm NS}\lesssim 0.68$ 
\citep{Most2020}. For this scenario to be viable the maximum mass of a spherical, nonrotating
cold NS has to be 
$\gtrsim 2.1\; M_\odot$ \citep{Most2020,Tsokaros2020}. Requiring rapid rotation for a NS companion
in GW190814 is a direct  consequence of the likely upper limits ($2.2-2.3\; M_\odot$) placed on
a spherical, nonrotating NS mass by event GW170817 
\citep{Margalit2017,Shibata2017,Rezzolla_2018,Ruiz:2017due,Shibata2019}. These upper limits were
mostly based on the assumption that the companions in GW170817 were slowly rotating. Assuming rapid
uniform NS rotation, instead, the upper limit allowed by event GW170817 increases~\citep{TheLIGOScientific:2017qsa,Ruiz:2017due}
and can explain the~$2.6\,M_\odot$~compact object in GW190814 as a slowly rotating
NS. In fact, by allowing for the uncertainties and adopting a sufficiently stiff  EOS, even  a nonrotating
NS can explain GW190814~\citep{Tsokaros2020}.
Note that although no robust discovery of a BHNS exists yet, the NSs in the 20 known NSNS systems 
\citep{Tauris:2017omb,Zhu:2017znf} have low dimensionless spins.  While one cannot draw definitive conclusions
from these limited number of observations, one might safely argue that if spin-down due to EM emission is as 
efficient as in currently known binaries, then any scenario involving a highly spinning NS 
either in an NSNS (like GW170817) or in an BHNS system (like GW190814) is not probable.
In summary, invoking rotation to explain the companion to the BH object in GW190814 depends on the stiffness
of the EOS and the assumptions of the maximum mass of a spherical NS. For a soft EOS (low spherical maximum
mass) such as SLy~\citep{Douchin01} rapid rotation 
is not sufficient, while for sufficiently stiff EOS such as DD2~\citep{HEMPEL2010210} rapid rotation may not even
be necessary. Such EOSs are neither rejected nor favored  by GW170817, and they are in accordance with the
results of NICER~(see~Fig.~\ref{fig:Tsokaros_gw19}). 
%
%%%%%%%%%%%%%%%%%%%%%%%%%%%%
%%%  Binary Neutron star %%%
%%%%%%%%%%%%%%%%%%%%%%%%%%%%
 \section{\bf NSNS mergers: Remnants and incipient jets}
 Numerical simulations of NSNS binaries are somewhat simpler than BHNS binaries, since the latter
 must treat the BH singularity. Some of the first numerical studies of NSNSs employed Newtonian
 gravity, modeling the NS as a polytrope~\citep{Gilden:1983hs,1989PThPh..82..535O,1992ApJ...401..226R,
   1994ApJ...432..242R,10.1143/ptp/88.6.1079,10.1143/ptp/89.4.809,Xing:1994ak,New_1997}. For
 circular orbit binaries it was found that following the binary merger, a highly differentially
 rotating remnant is formed. However, their simulations could not track its possible collapse to a BH
 with Newtonian gravity.
 Motivated by models of sGRBs and the  ejection of r-process nuclei,
 \cite{1994ApJ...431..742D},~\cite{1996A&A...311..532R} and~\cite{1998A&A...338..535R} extended
 the previous results by incorporating a simple treatment of the nuclear physics in their numerical
 calculations.  One of the first approaches used to simulate NSNS coalescence in GR
 was the ``conformal flatness approximation'' (CFA) used by~\cite{1995PhRvL..75.4161W}, which has been
 followed by several other treatments with increasing sophistication. \cite{Oechslin:2001km} 
 evolved NSNS binaries using a Lagrangian SPH code with a multigrid
 elliptic solver to handle the gravitational field equations and corotating initial configurations.
 \cite{Faber:2003sb} subsequently performed SPH simulations in the CFA using a spectral elliptic
 solver in spherical coordinates and  employed the quasi-equilibrium, irrotational binary models of
 \cite{tg02}. These models are constructed using the conformal thin-sandwich formalism~\citep{1999PhRvL..82.1350Y}.
 \cite{2007A&A...467..395O} extended their earlier studies by including the influence of a realistic nuclear EOS.
 These simulations showed that the dynamics and the final outcome of the merger depend sensitively
 on the EOS and the binary parameters, such as the gravitational mass of the system and its mass ratio. 
 The first fully GR simulations of NSNS undergoing  merger
 were performed by~\cite{ShiU00,Shibata:2002jb} and \cite{STU1} using a polytropic EOS to model the stars.
 Since then, great progress has been made to model NSNSs incorporating realistic microphysics and
 magnetic field effects in full GR and in alternative theories of gravity. In the following we only
 review full GR studies of these binaries. For earlier reviews and references, see, e.g.,
 \cite{BSBook} and~\cite{shibatabook}.

\vspace{0.5cm}
%%%%%%%%%%%%%%%%%%%%%%%%%%%%%%%%%%%
%%%   Nonmagnetized evolutions  %%%
%%%%%%%%%%%%%%%%%%%%%%%%%%%%%%%%%%%
\subsection{\bf Nonmagnetized evolutions}
\label{sec:nsns_nonmag}
One of the first questions numerical studies of NSNS mergers in full GR were compelled to
address was under what conditions the highly differentially rotating star remnant collapses
to a BH. The uncertainties in the nuclear EOS, combined with theoretical arguments  invoking
GW170817 and its EM counterparts, allow nonrotating NSs with a %EOSs with the so-called Tolman–Oppenheimer–Volkoff
%(TOV)
maximum mass limit in the range $\maxtov\sim 2.1-2.4\,M_\odot$~\citep{Margalit2017,Rezzolla_2018,
  Ruiz:2017due,Shibata2017,Shibata2019}.
%See also~\citep{Bauswein:2013jpa} for realistic finite-temperature EOSs. 
Uniform rotation allows NSs with up to~$\sim 20\%$ more mass
(``supramassive stars''; as coined by~\cite{Cook:1993qr,Cook:1993qj}). Even
larger masses can be supported
against collapse with centrifugal support if the star is differentially rotating. Such  stars
were first constructed and explored by~\cite{Baumgarte:1999cq}, who built dynamically stable $\Gamma=2$
polytropic models with masses $\gtrsim 3-4\,M_\odot$. They coined the label ``hypermassive neutron star'' (HMNS)
to describe such stars. It was demonstrated by~\cite{Duez:2004nf} that shear viscosity drives
a HMNS to collapse to a BH on a (secular) viscous timescale and by~\cite{dlsss06a} that turbulent magnetic
viscosity induced by MRI can also drive the secular collapse of the latter magnetic HMNSs. These viscous
effects compete with neutrino  and GW emission (when the HMNS remnant is nonaxisymmetric)
to drive collapse. In NSNS binaries, the fate of the remnant depends on the total mass of the
NSNS binary, as we shall now discuss.

\cite{ShiU00} and \cite{Shibata:2006nm} found that there is a threshold mass $M_{\rm th}$
above which the remnant collapses immediately on a dynamical timescale to a BH, independently of the
initial binary mass ratio. This threshold value depends strongly on the EOS. For $\Gamma=2$ polytropes
$M_{\rm th}\approx 1.7\,\maxtov$, while for stiffer EOSs, such as APR~\citep{Akmal:1998cf} and
SLy~\citep{Douchin01}, it is $\sim 1.3-1.35\,\maxtov$.
\cite{Shibata:2006nm} also found that in the case of ``prompt'' collapse to a BH, the  mass of the
disk remnant increases sharply with increasing mass ratio for a fixed gravitational mass and EOS. In
addition, if the mass of the binary is less than $M_{\rm th}$ the disk remnant
turns out to be more massive than for those whose mass is larger than $M_{\rm th}$. For binaries with
$M<M_{\rm th}$ their remnants form a transient, highly deformed HMNS which, after
$\sim 8-50\rm\,ms$, undergoes a ``delayed'' collapse to a BH
surrounded by a significant accretion disk. The collapse occurs due to angular momentum losses from
gravitational radiation in these simulations where neutrino cooling and magnetic fields are absent
\citep{Baiotti:2008ra,Kiuchi:2009jt,Rezzolla:2010fd,Dietrich2015,Ruiz:2019ezy}. 
These results have been extended by~\cite{Hotokezaka:2011dh} using a piecewise polytropic representation
of nuclear EOSs~\citep{Read:2008iy,2009PhRvD..80j3003O}.  It was found that the threshold value
is in the range $1.3 \lesssim M_{\rm th}/\maxtov \lesssim 1.7$. 
%A long-lived HMNS is formed
%for binaries with total mass of $\sim 2.7\,M_\odot$~and for the EOS for which $\maxtov$ exceeds
%$2.0\,M_\odot$, such as APR4~\citep{Akmal:1998cf} and H4~\citep{PhysRevLett.67.2414}.
These results were
confirmed  also for realistic finite-temperature
EOSs~\citep{Bauswein:2013jpa}. In addition, the ratio between the
threshold mass and maximum mass is tightly correlated with the compactness of the~$\maxtov$.
Finally, less massive binary mergers  form a
dynamically stable NS remnant  that may collapse on
longer time scales once dissipative processes, such as neutrino dissipation or gravitational radiation,
take place~\citep{Cook:1993qj,Cook:1993qr,1996ApJ...456..300L,Breu:2016ufb}.

Most of the numerical calculations to date have focused on quasi-circular irrotational binaries,
though it is expected that spin can modify the threshold value of prompt collapse, or at least
change the lifetime of the remnant. Preliminary results reported by~\cite{Kastaun2015},
\cite{Dietrich:2016lyp}, \cite{Ruiz:2019ezy} and \cite{Chaurasia:2020ntk} showed that  depending on the
NS spin, the lifetime of the remnant may change from $\sim 8$ to $\gtrsim 40\,\rm ms$. Effects of NS
spin on the inspiral have been explored by~\cite{PhysRevD.96.084060}, \cite{Bernuzzi:2013rza},
\cite{Dietrich:2017xqb} and \cite{Tsokaros:2019anx}.
 On the other hand, the dynamically captured NSNS mergers that  may arise in dense stellar regions,
such as globular clusters, have been studied by~\cite{East2012NSNS}. These results showed that
$M_{\rm th}$ and the mass of the disk remnant depend not only on the EOS but also on the impact
parameter. The calculations by~\cite{Paschalidis:2015mla} and~\cite{PEFS2016} demonstrated that the
HMNS formed through dynamical capture may undergo the one-arm nonaxisymmetric (mode $m=1$)
instability.

During merger, shock heating produces temperatures as high as  $\sim 100\,\rm MeV$ at the contact
layer between the two stars. Subsequent compressions lead to average-temperatures of the order of
$10\,\rm MeV$ in the central core of the NSNS remnant~\citep{Bauswein:2010dn}, and hence the binary
remnant can be a strong emitter of neutrinos. The timescale of neutrino cooling radiation (typically
$\lesssim 1\,\rm s$) may also strongly affect the HMNS lifetime~\citep{Sekiguchi:2011zd}. Effects of
neutrino  cooling on the dynamical ejecta that can give rise to
observable kilonova signatures have been studied in~\cite{Radice:2016dwd},
\cite{Lehner:2016lxy}, and~\cite{Sekiguchi:2015dma} (see~\cite{Radice:2020ddv} for a recently review).
It is worth noting that the calculations of~\cite{Bauswein:2010dn} and~\cite{Just:2015dba} show that
neutrino heating drives a wind from the surface of the remnant, creating very baryon-loaded environments
in the polar regions that prevent the formation of incipient jets. Therefore, MHD processes are likely be a
key ingredient to overcome this and to trigger the formation of relativistic jets.

%%%%%%%%%%%%%%%%%%%%%%
%%% binary P=rho  %%%%
%%%%%%%%%%%%%%%%%%%%%%
\begin{figure}
  \begin{center}
\includegraphics[width=8.6cm]{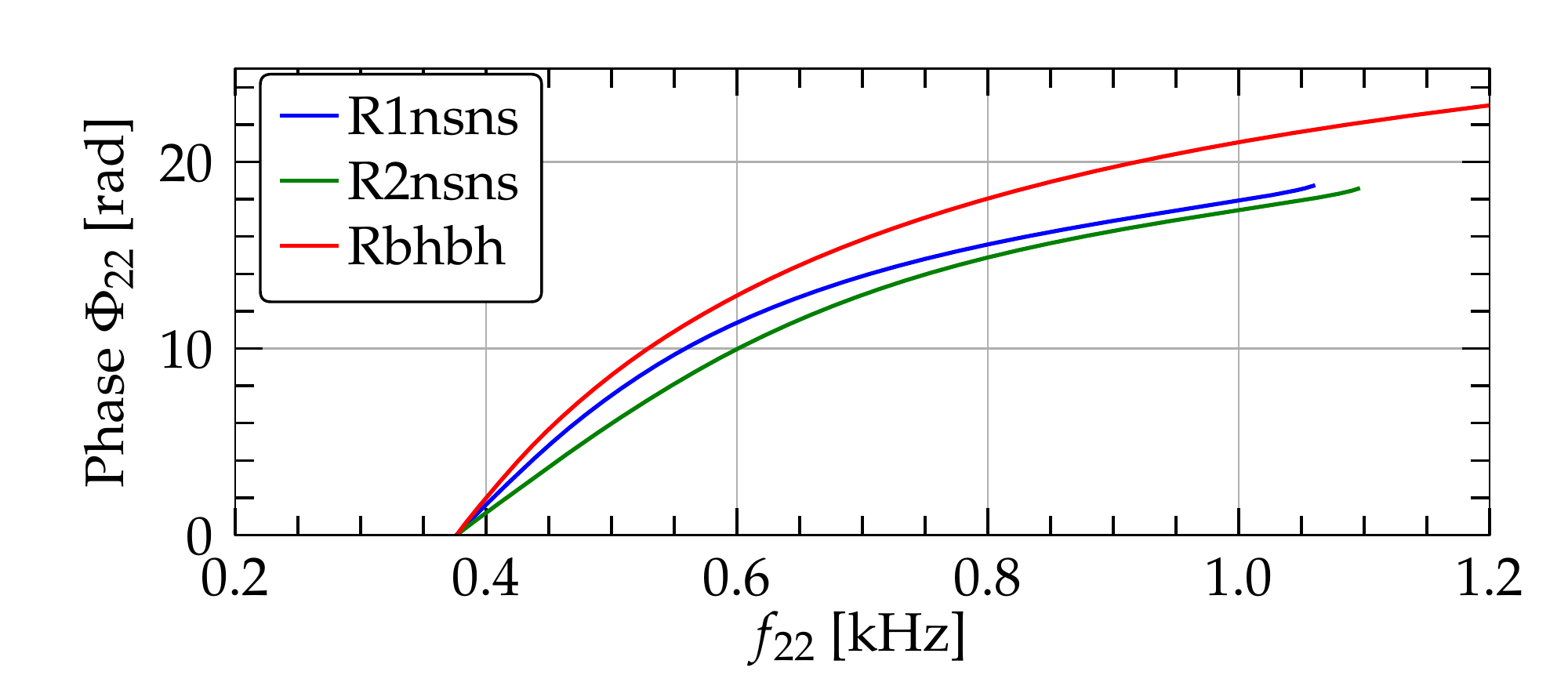} 
\includegraphics[width=8.6cm]{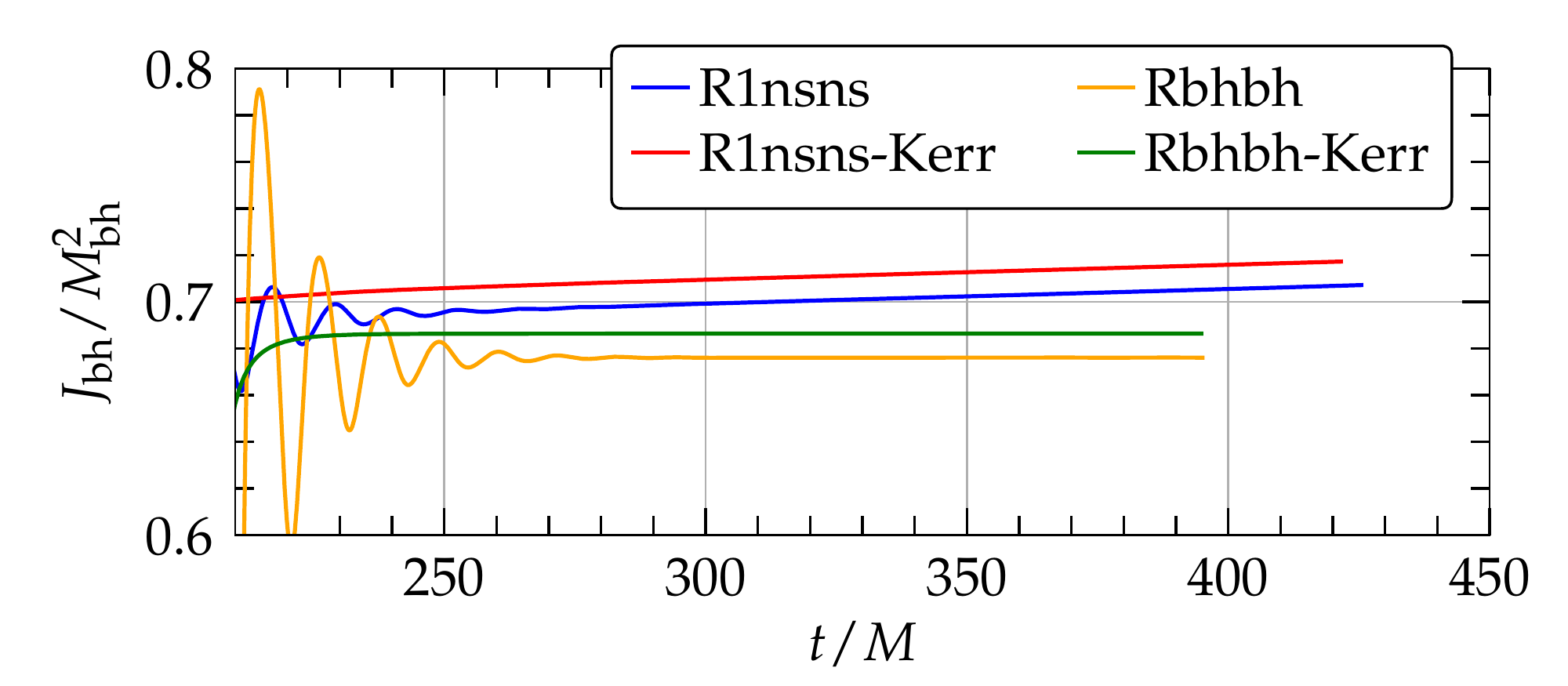}
\caption{Left panel: GW phase versus frequency for the NSNS binary using two resolutions (R1nsns,R2nsnsn)
and a BHBH binary having the same gravitational mass. Right panel: Dimensionless spin of the remnant BH
for the NSNS (R1nsnsn) and the BHBH (Rbhbh) binary. Also shown is the dimensionless spin as computed 
from the Kerr formula for the two systems [From~\cite{Tsokaros:2019lnx}].}  
%\caption{Gravitational mass  vs rest-mass  density for the ALF2 and
%  ALF2cc EOSs for equilibrium models in spherical symmetry (left), and
%  rest-mass density profile along the x-axis for the NSNS binary
%  with the ALF2cc EOS (right). Horizontal red line corresponds to nuclear
%  density $\rho_{0,\rm nuc}$.   Insets show the NS compaction vs rest-mass
%  density (left), and a zoom into the area close to the NS surface
%  where the density steeply drops from $\rho_{0,\rm nuc}$ to zero (right) [adapted from Figs.
%  1 and 5 in~\citep{Tsokaros:2019lnx}].}  
\label{fig:rho0}                                                                 
\end{center}                                                                     
\end{figure} 

Numerical simulations of NSs having a mass that falls inside the so-called ``mass-gap'' are scarce. The 
first such simulation was performed by~\cite{Tsokaros:2019lnx} with a binary NSNS in a quasi-equilibrium 
circular orbit. The gravitational mass of the binary
was $M=7.90\,M_\odot$, and each star is identical and  has  a compactness of $\mathcal{C}=0.336$. This value,
which is even higher than the maximum possible compactness that can be achieved by solitonic boson
stars~\citep{Palenzuela:2017kcg},
is slightly smaller than the limiting compactness $\mathcal{C}_{\rm max}=0.355$ set by causality~\citep{2016PhR...621..127L}.
To build these binaries,~\cite{Tsokaros:2019lnx} employed the ALF2 EOS~\citep{Alford:2004pf}, but replaced the
region where the rest-mass density satisfies~$\GR_0\geq\GR_{0,s}=\GR_{0,\rm nuc}=2.7\times 10^{14}\ {\rm gr/cm^3}$
by the maximally stiff EOS 

\begin{equation}
P=\GR-\GR_s + P_s \,,
\label{eq:eoscc}
\end{equation}
with sound speed equal to the speed of light.
Here $\GR$ is the total energy density, and $P_s$ the pressure at $\GR_s$, assumed known. 
The quasi-equilibrium initial data were built using the COCAL code~(see e.g.~\cite{Tsokaros:2015fea, Tsokaros:2016eik}). 
%Spherical equilibrium models for both the ALF2 and the resulting EOS (denoted as
%ALF2cc) are displayed on~left panel in~Fig.~\ref{fig:rho0} where the gravitational mass $M$ and the
%compactness of the stars (inset) are plotted vs the rest-mass density. 
%For the ALF2 EOS the maximum mass is $M_{\rm max}^{\rm sph}=2.0M_\odot$ at rest-mass density $1.63\times 10^{15}\ {\rm gr/cm^3}$
%and compactness $0.26$, while for the ALF2cc EOS we have  $M_{\rm max}^{\rm sph}=4.06M_\odot$
%at $\GR_0=6.0\times 10^{14}\ {\rm gr/cm^3}$ and compactness $0.349$. 
%Right panel shows the rest-mass density profile along the x-axis which passes through the centroid of each star.  The profiles
%resemble those of self-bound quark stars whose density at the surface is finite.  These results
%found that these extreme compact binaries undergoing prompt collapse to a BH. The ringdown remnant
%BH is consistent with the ringdown of a perturbed Kerr BH.
%with the mass and angular momentum of
%the remnants  closely matching the ones predicted by the Kerr metric.
Due to the large compactions of the NSs the binary stars exhibit no tidal disruption up until
merger, whereupon a prompt collapse is initiated even before a common core forms.
Within the accuracy of the simulations the BH remnant from this NSNS binary 
exhibits ringdown radiation that is not easily distinguishable from a perturbed
Kerr BH. Right panel of Fig.~\ref{fig:rho0} displays the dimensionless spin from the 
BH remnant from the NSNS  and that from a BHBH binary  having the same gravitational (ADM) mass.
Also shown are the remnant spins
as computed from the analytic Kerr formula whose input is the ratio $L_p/L_e$ of the polar to 
equatorial circumference.
However, the inspiral  leads to phase differences of the order
of $\sim 5$ rad (left panel of Fig.~\ref{fig:rho0}) over an $\sim 81$ km separation ($\sim 1.7$ orbits). 
Although such a difference can be measured by current GW laser interferometers 
(e.g. LV scientific collaboration observatories), uncertainties in the individual masses and spins will likely 
prevent distinguishing such compact, massive NSNSs from BHBHs.

\vspace{0.5cm}
%%%%%%%%%%%%%%%%%%%%%%%%%%%%%
%%% Magnetized Evolutions %%%
%%%%%%%%%%%%%%%%%%%%%%%%%%%%%
\subsection{\bf Magnetized evolutions}
Although NS may have very large magnetic fields ($\gtrsim 10^{14}\,\rm G$) at birth, it is expected
that cooling processes significantly reduce their magnitudes~\citep{Pons:2008fd}. Pulsar observations
indicate that the characteristic surface magnetic field strength of NSs~is~$\sim 10^{10}-10^{12}\,\rm
G$ \citep{1990puas.book.....L,lrr-2008-8,Miller_2019,Semena:2019juf}.  Nevertheless, magnetic instabilities
such as the Kelvin-Helmholtz instability~(KHI; see e.g.~\cite{Price:2006fi,Anderson:2008zp,Kiuchi:2015qua,
  Kiuchi:2015sga}), MRI~(see~e.g.~\cite{dlsss06a,Kiuchi:2015qua,Shibata:2005mz,Siegel:2013nrw}),
and magnetic winding (see~e.g.~\cite{Baumgarte:1999cq,Kiuchi:2015sga,Sun:2018gcl}) triggered during and
after the NSNS merger can substantially boost the strength of these weak fields.

%%%%%%%%%%%%%%%%%%%%%%
%%% Aguilera BNS  %%%%
%%%%%%%%%%%%%%%%%%%%%%
\begin{figure}                                                                   
\begin{center}                                                                   
\includegraphics[width=18.6cm]{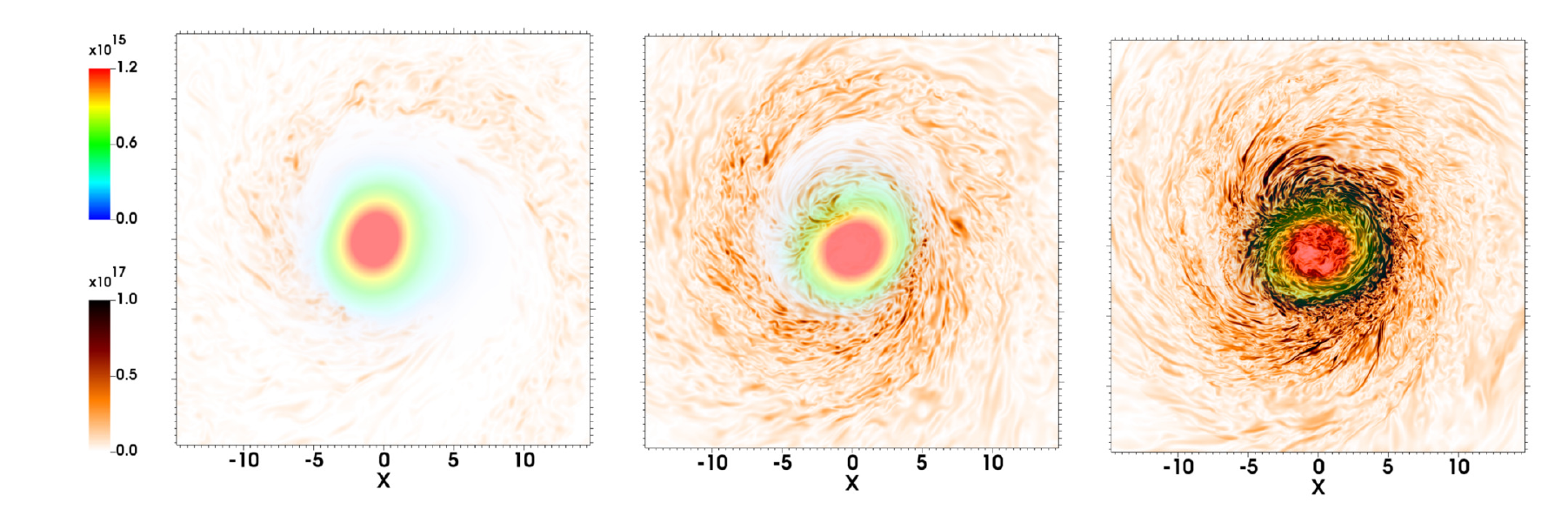}                            
\caption{NS rest-mass density $\rho_0$ (upper left rainbow colorbar) and magnetic field (lower left brownish colorbar)
  on the orbital plane at $t=10.0\,\rm ms$  following an NSNS merger at three different resolutions
  [$\Delta x=147\rm\,m$ (left), at $\Delta x=74\rm\,m$ (middle), and $\Delta x=37\rm\,m$ (right)].
  The initial magnetic field strength is $5\times 10^{11}\,\rm G$ [Adapted from~\cite{Aguilera-Miret:2020dhz}].}
\label{fig:Magnetic_growth}
\end{center}                                                                     
\end{figure} 

High-resolution simulations are required to properly capture the above instabilities because their
fastest growing modes have short wavelengths. \cite{Kiuchi:2015qua,Kiuchi:2015sga} systematically
studied the effects of numerical resolution on the magnetic field amplification in NSNS mergers
and found that, at the unprecedented resolution of $\Delta x=17.5\,\rm m$, an initial magnetic field
of $10^{13}\,\rm G$ is amplified to values $\gtrsim 10^{15} \,\rm G$ in the bulk of the remnant,
with local values peaking at $\sim 10^{17}\,\rm G$, after $5\,\rm ms$ following merger. Recently,
the calculations by~\cite{Aguilera-Miret:2020dhz} reported that at a resolution of~$\Delta x=37\,\rm m$
an initial magnetic field of $5\times 10^{11}\,\rm G$ is amplified to values of $\sim 10^{17}\,\rm G$
after about $10\,\rm ms$ following merger (see~Fig.~\ref{fig:Magnetic_growth}).  These extremely
high-resolution simulations  are computationally quite expensive and currently inaccessible for general
studies. Typical NSNS simulations use a resolution~$\gtrsim 120\,\rm m$~(see~e.g.
\cite{Ciolfi2019,Ruiz:2019ezy,Weih:2019xvw,Vincent:2019kor,Bernuzzi:2020txg}). To overcome
the lack of resolution,  some works have adopted subgrid models to mimic the effect of
magnetic instabilities~(see~.e.g.~\cite{Giacomazzo:2014qba,Palenzuela:2015dqa,Aguilera-Miret:2020dhz,
  Radice:2020ids}), while others have employed high, but dynamically weak initial magnetic
fields to mimic the resulting magnetic field following the merger~(see~e.g.~\cite{Ruiz:2016rai,
  Ciolfi2019,Mosta:2020hlh}). These two approaches allow the tracking of the secular evolution of the
a quasi-stationary NSNS remnant consisting of a HMNS that ultimately undergoes delayed collapse to
a highly spinning BH surrounded by an accretion disk with a strong magnetic field with finite
computational resources.
%
%%%%%%%%%%%%%%%%%%%%
%%%  Kiuchi BNS  %%%
%%%%%%%%%%%%%%%%%%%%
%
\begin{figure}                                                                   
\begin{center}                                      
\includegraphics[width=18.2cm]{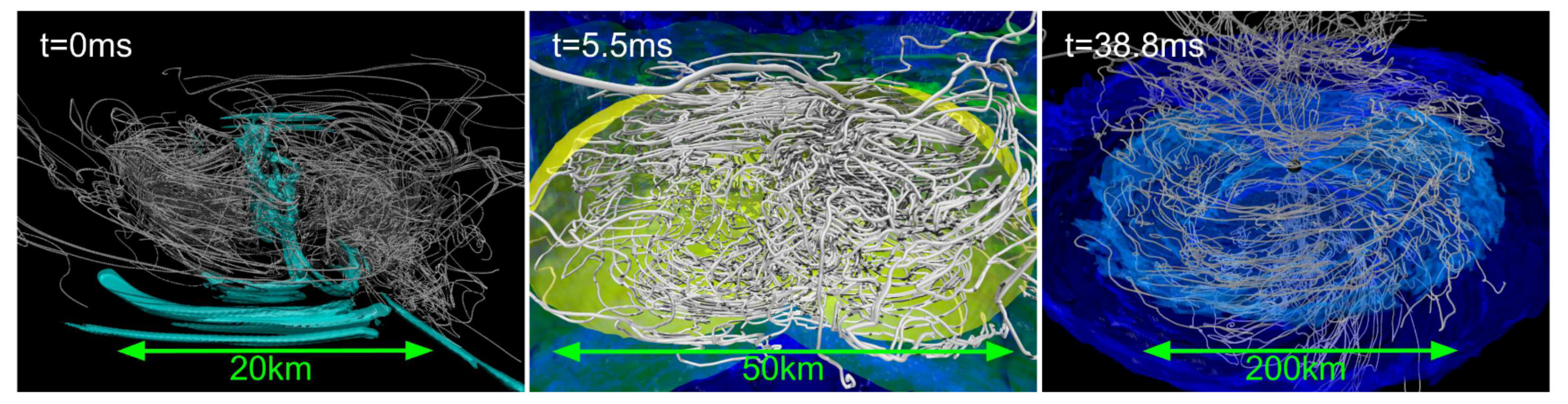}         
\caption{NS rest-mass density and magnetic field lines at $t-t_{\rm mrg}\approx 0.0\,\rm ms$ (left),
  $t-t_{\rm mrg}\approx 5.5\,\rm ms$ (middle), and $t-t_{\rm mrg}\approx 38.8\,\rm ms$ (right)
  following a NSNS merger.
  Here $t_{\rm mrg}$ is the merger time. Cyan color on the left panel displays  magnetic fields
  stronger than $10^{15.6}\,\rm G$. Yellow, green, and dark blue colors on  the middle panel
  show rest-mass densities of $10^{14}$, $10^{12}$, and $10^{10}\,\rm g/cm^3$, respectively.
  Light and dark blue colors on the right panel indicate rest-mass densities of $10^{10.5}$, and
  $10^{10}\,\rm g/cm^3$, respectively  [From~\cite{Kiuchi:2015qua}].}
\label{fig:H4_BNS}                                                                 
\end{center}                                                                     
\end{figure} 

Some of the first long-term ideal GRMHD studies of NSNS mergers were performed by~\cite{Anderson:2008zp}
and~\cite{lset08} using $\Gamma=2$ polytropes endowed with a $10^{16}\rm\,G$ polodial magnetic field
confined to the NS interior (see Eq.~\ref{ini:Aphi_int}). The simulations of~\cite{Anderson:2008zp} reported
the formation of a long-lived HMNS.  During this phase, turbulent magnetic fields transport angular
momentum away from the center, inducing the formation an axisymmetric central core that eventually
collapses to a spinning BH.~\cite{lset08} reported  the evolution of equal and unequal binaries that
promptly collapse to a BH following merger, surrounded by a disk with $\lesssim 2\%$ of the total
rest mass of the binary.  Neither  an outflow nor a magnetic field collimation were found.

The calculations of~\cite{rgbgka11} reported that $\sim 12\,\rm ms$ after the collapse of a
HMNS remnant, MHD instabilities develop and form a  central,  low-density,  poloidal-field
funnel, though there were no evidences of an outflow. The initial data consist of a binary
polytrope initially endowed with a $10^{12}\,\rm G$ poloidal magnetic field confined to the
stellar interior. The highest resolution used in these studies was~$\Delta x\approx 221\,
\rm m$.  A subsequent high-resolution study by~\cite{Kiuchi:2015qua}, employing an H4 EOS
\citep{PhysRevLett.67.2414} with seed poloidal magnetic fields confined to the stellar interior,
found that during merger, the magnetic field is steeply amplified due to the KHI. In their
high-resolution case ($\Delta x=70\,\rm m$) the amplification is $40-50$ times larger than that
in the low-resolution case~($\Delta x=150\,\rm m$).  In contrast to~the results of~\cite{rgbgka11},
the ram pressure of the fall-back
debris prevents the formation of a coherent poloidal field. As the frozen-in magnetic field lines are
anchored to the  fluid elements, an outflow, which was not seen after $40\rm\, ms$ following merger
(see Fig.~\ref{fig:H4_BNS}), is presumably necessary to generate a coherent poloidal magnetic field.

\begin{figure}                                                                   
  \begin{center}
    \includegraphics[width=5.9cm]{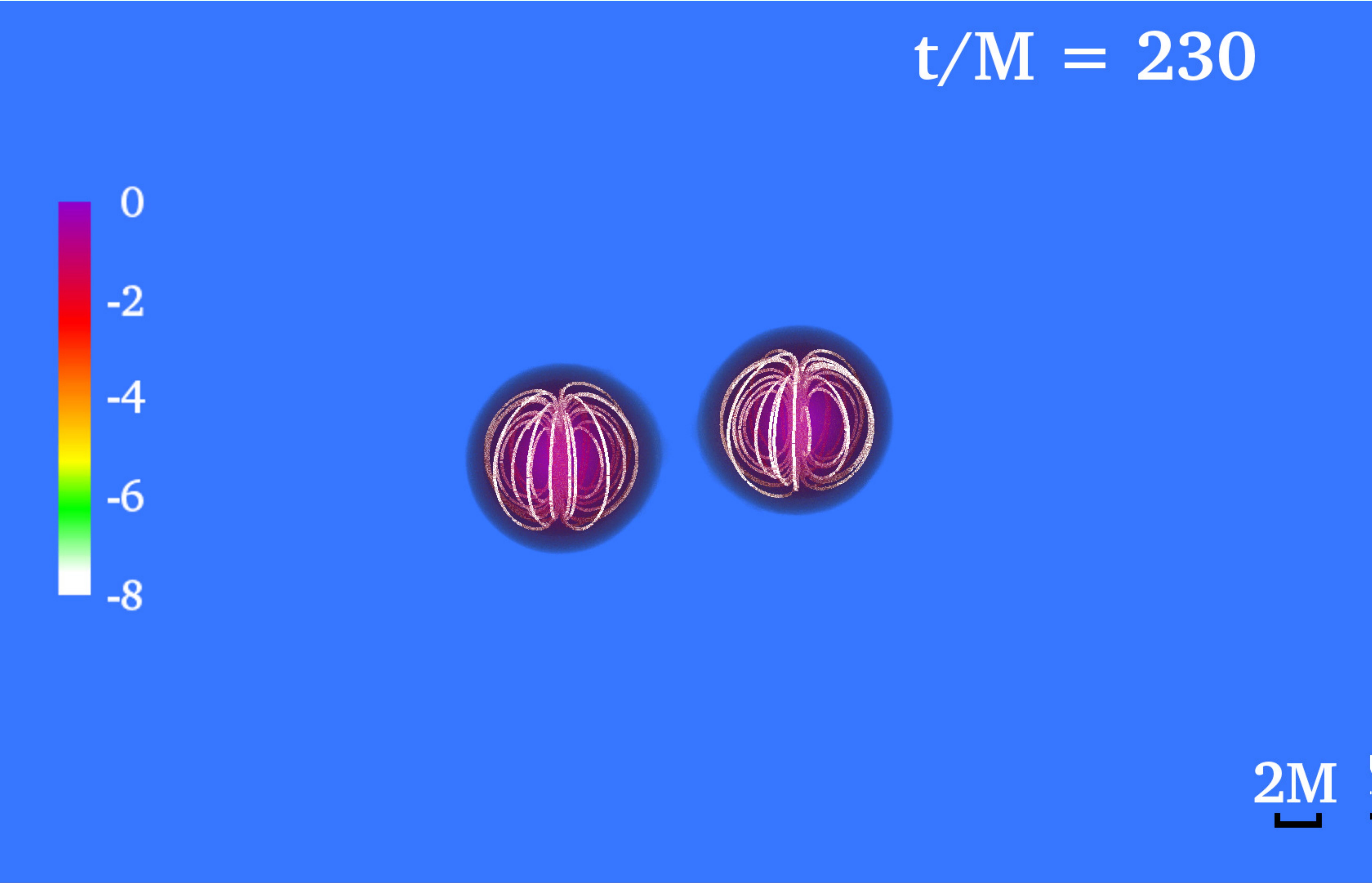}
    \includegraphics[width=5.9cm]{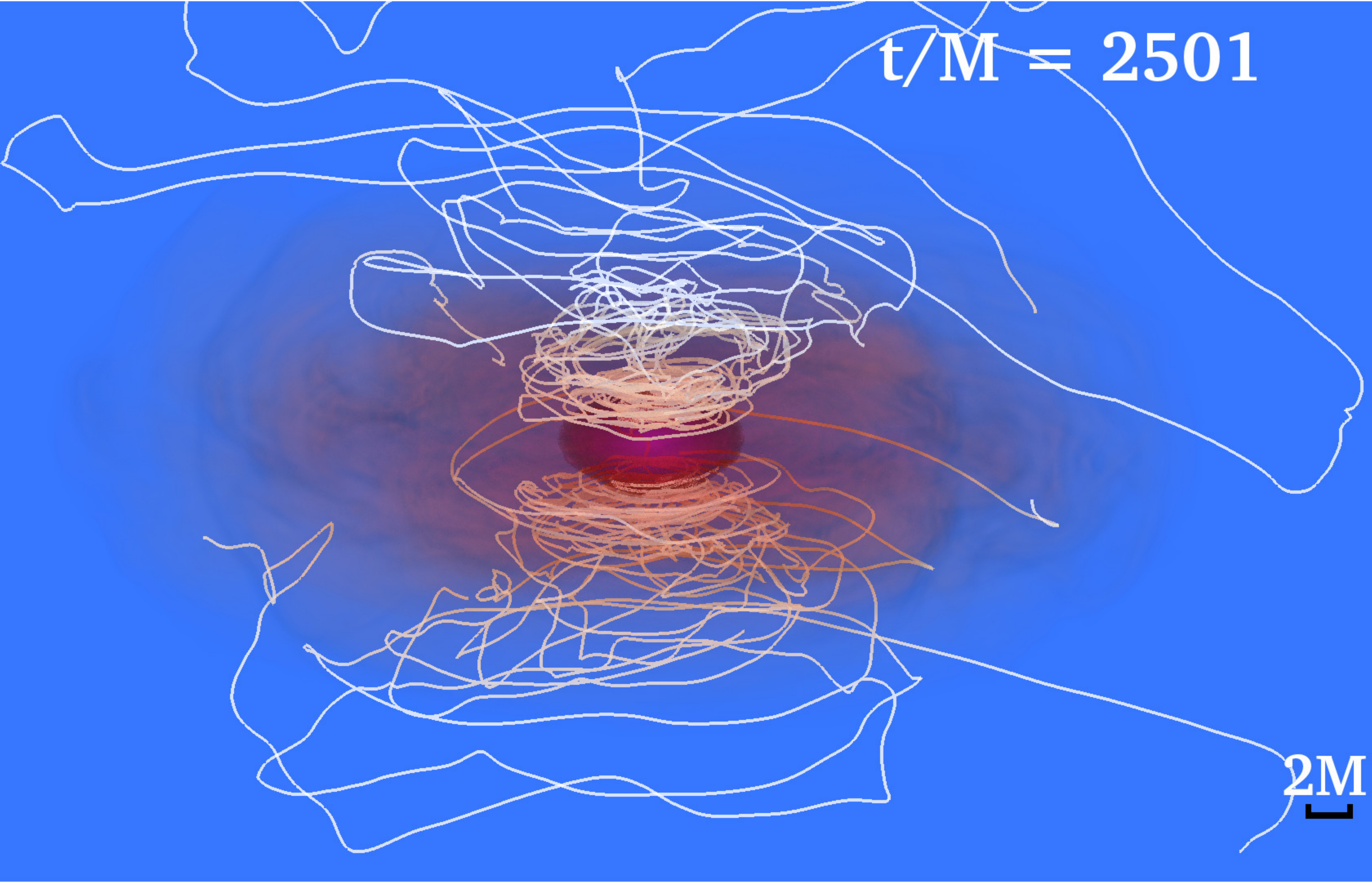}
    \includegraphics[width=5.9cm]{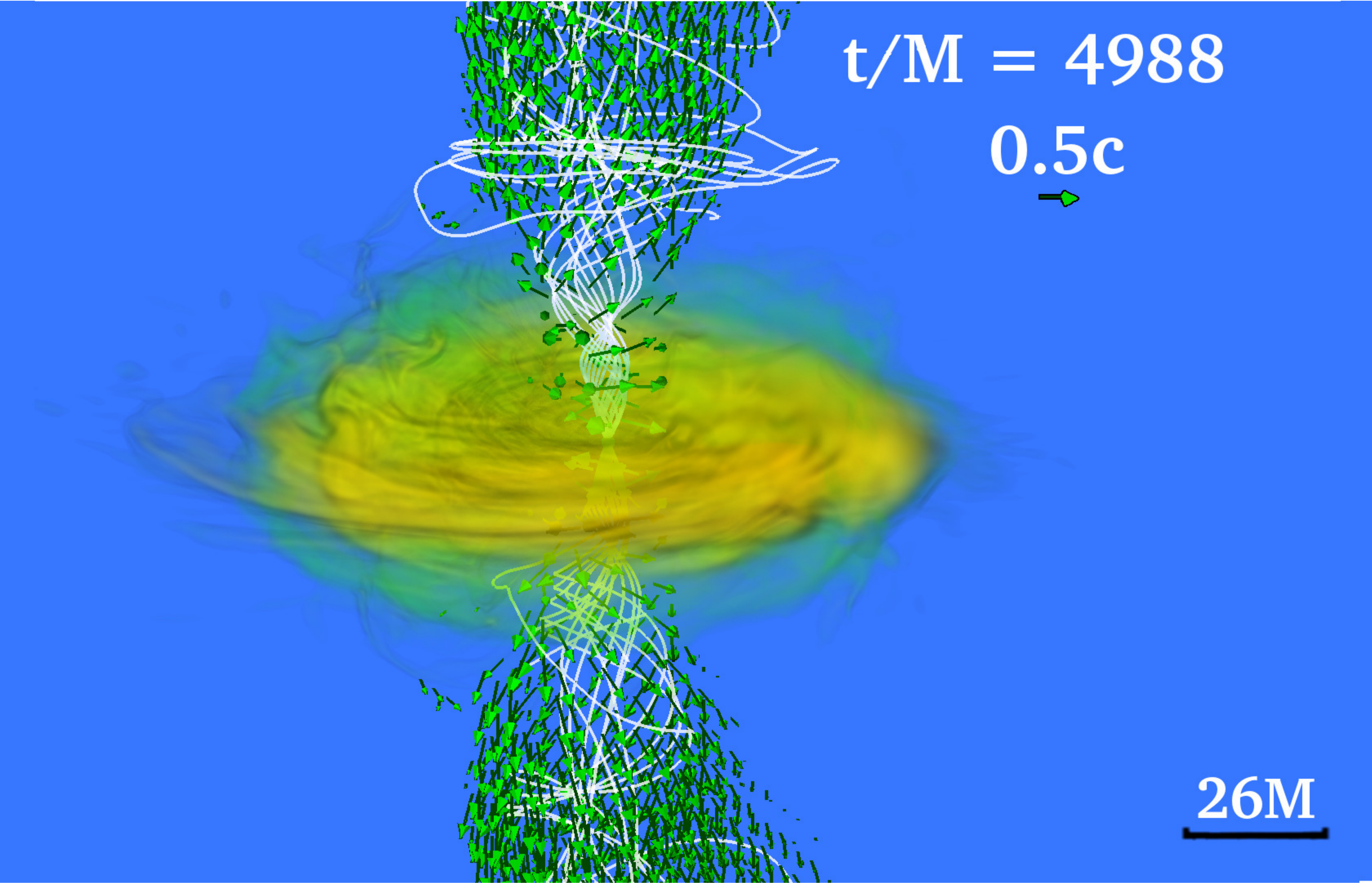}
    \caption{NS rest-mass density $\rho_0$ normalized to its
      initial maximum value $\rho_{0,\text{max}}= 5.9\times
      10^{14}(1.625\,M_\odot/M_{\rm NS})^2\,\text{g/cm}^3$ (log scale) at
      selected times for an NSNS merger.  Arrows display plasma velocities
      and white lines show magnetic field lines.  Here $M=1.47\times
      10^{-2}(M_{\rm NS}/1.625M_\odot)\,\rm ms$ = $4.43(M_{\rm
    NS}/1.625M_\odot)\,\rm km$ [Snapshots from case IH in~\cite{Ruiz:2016rai}].}
    \label{fig:BNS_ruiz}
  \end{center}                                                                     
\end{figure}

\cite{Ruiz:2016rai} evolved the same NSNS configuration as in~\cite{rgbgka11} but using higher
resolution ($\Delta x=152\,\rm m$). As this resolution is still not enough to properly capture the
growth of the magnetic field due to the KHI, \cite{Ruiz:2016rai} endowed the initial NSs with 
dynamically weak, purely poloidal magnetic fields with strengths ${B}_{\rm pole}\simeq 1.75\times
10^{15}(1.625M_\odot/M_{\rm NS})\,\rm G$ at the poles of the stars, which  matches the values of the field
strength in the HMNS reached in~\cite{Kiuchi:2015qua}. It was found that by $\sim 4000{\rm M}\sim
60 (M_{\rm NS}/1.625M_\odot)\rm\,ms$ following BH formation, the magnetic field above the BH poles
has been wound into a tight, helical funnel inside of which fluid elements begin to flow outward: this is
an incipient jet (see Fig.~\ref{fig:BNS_ruiz}). The lack of a jet in~\cite{Kiuchi:2015qua}  can be
attributed to the persistent fall-back debris in the atmosphere, which increases the ram
pressure above the BH poles. Therefore, a longer simulation like the one in~\cite{Ruiz:2016rai}  is 
required for jet launching. Notice that jet launching may
not be possible for all EOSs if the matter fall-back timescale is
longer than the disk accretion timescale~\citep{Paschalidis:2016agf}.

\begin{figure}
  \centering
  \includegraphics[width=5.9cm]{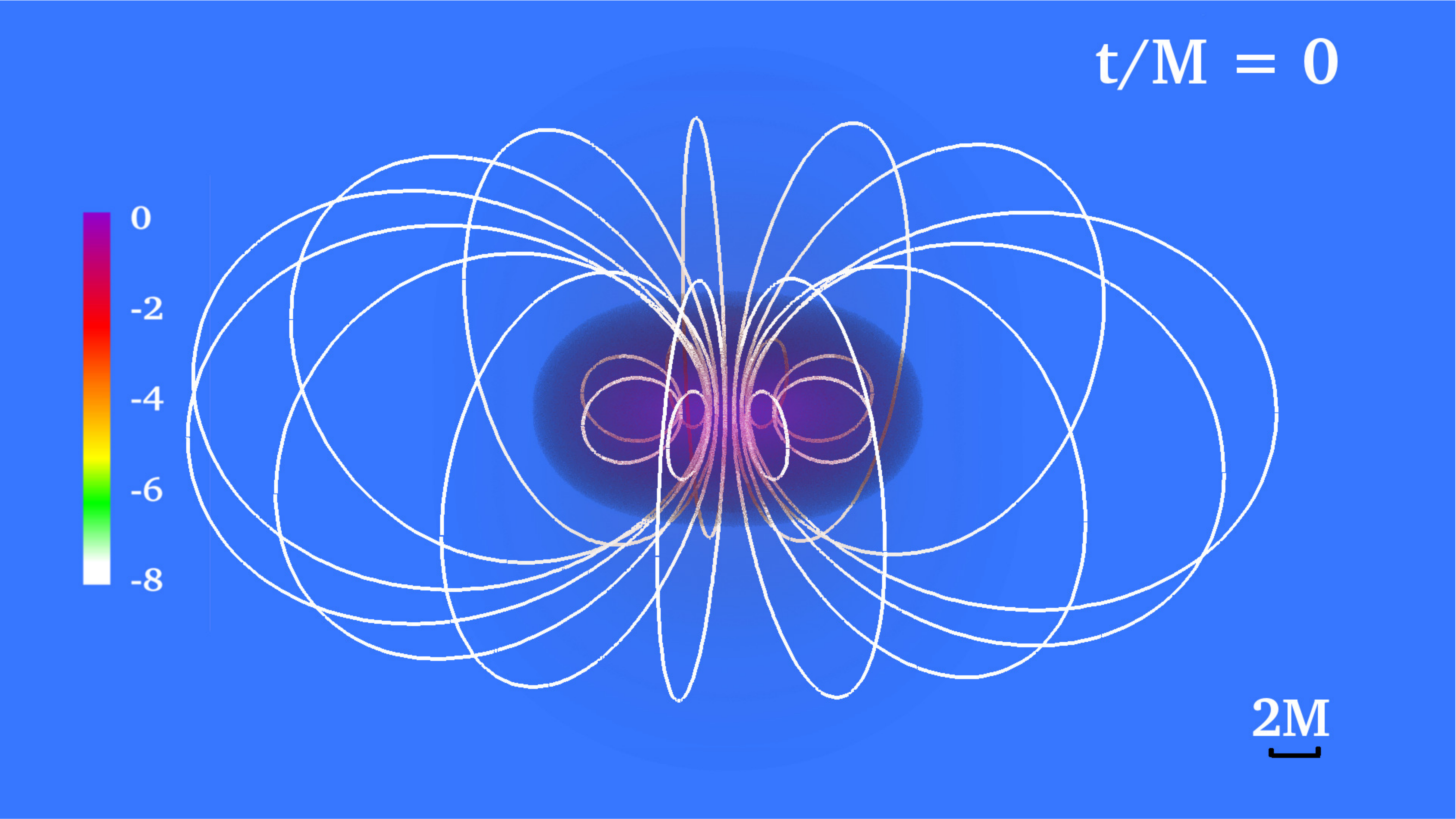}
  \includegraphics[width=5.9cm]{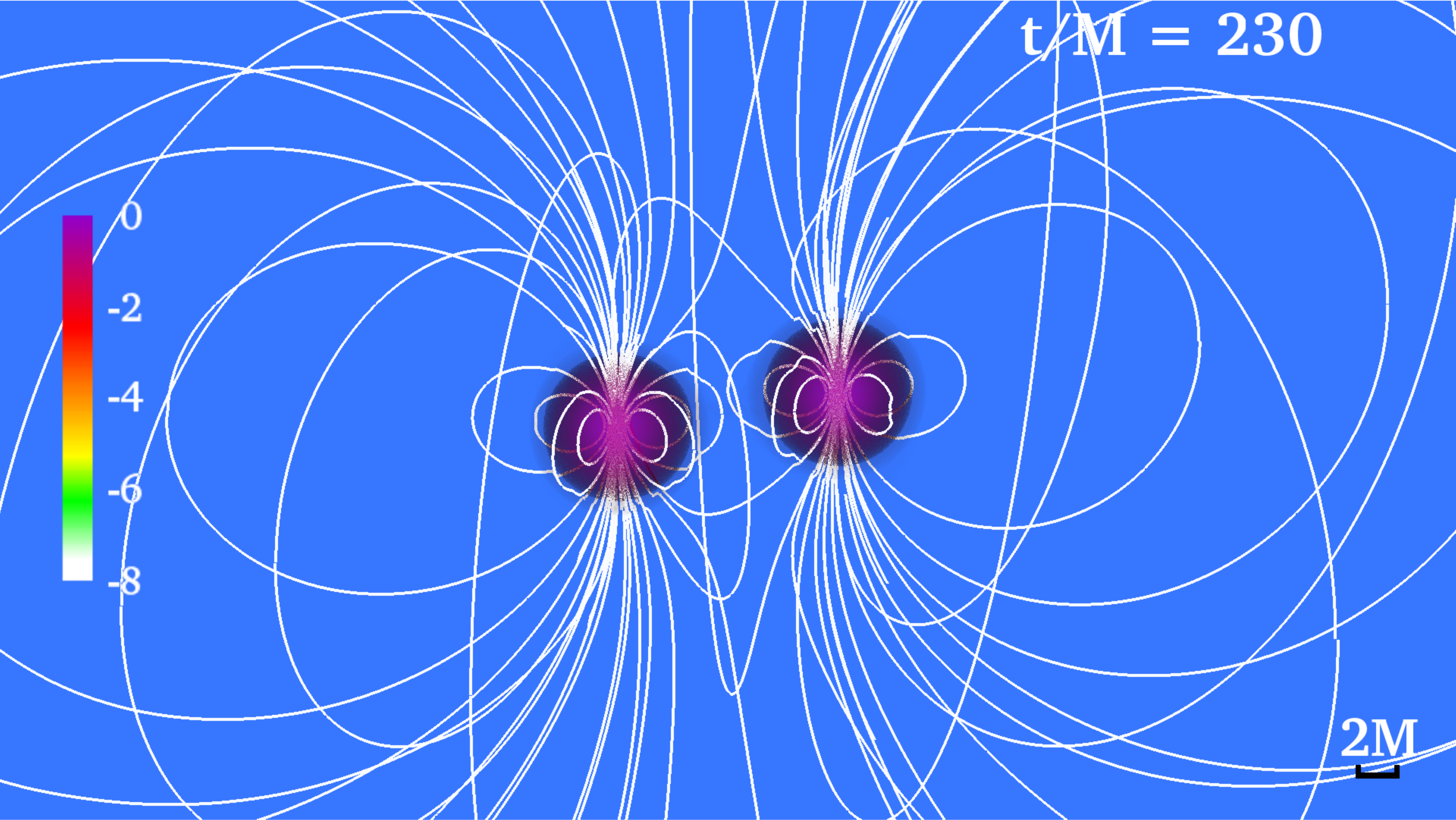}
  \includegraphics[width=5.9cm]{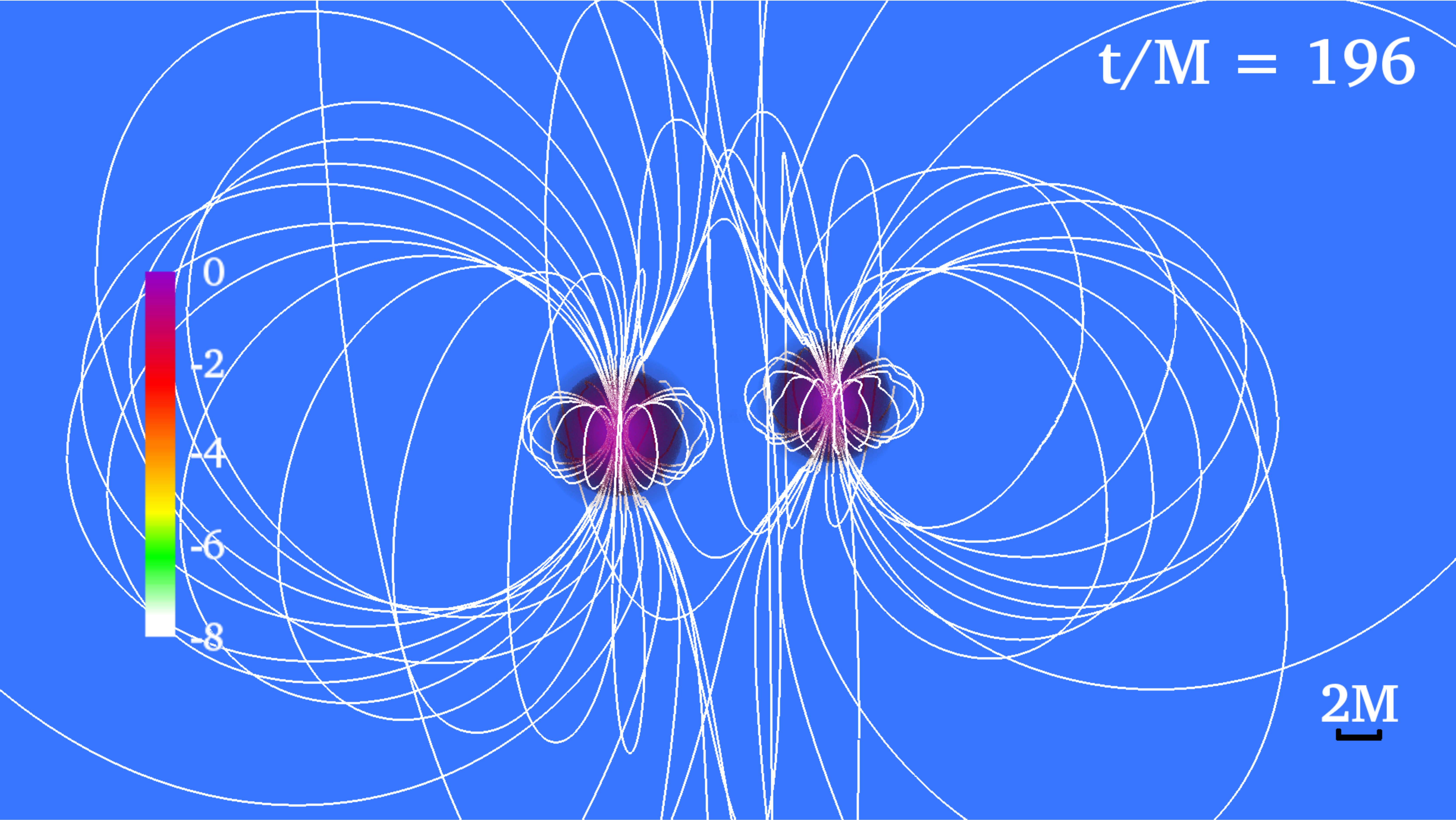}
  \includegraphics[width=5.9cm]{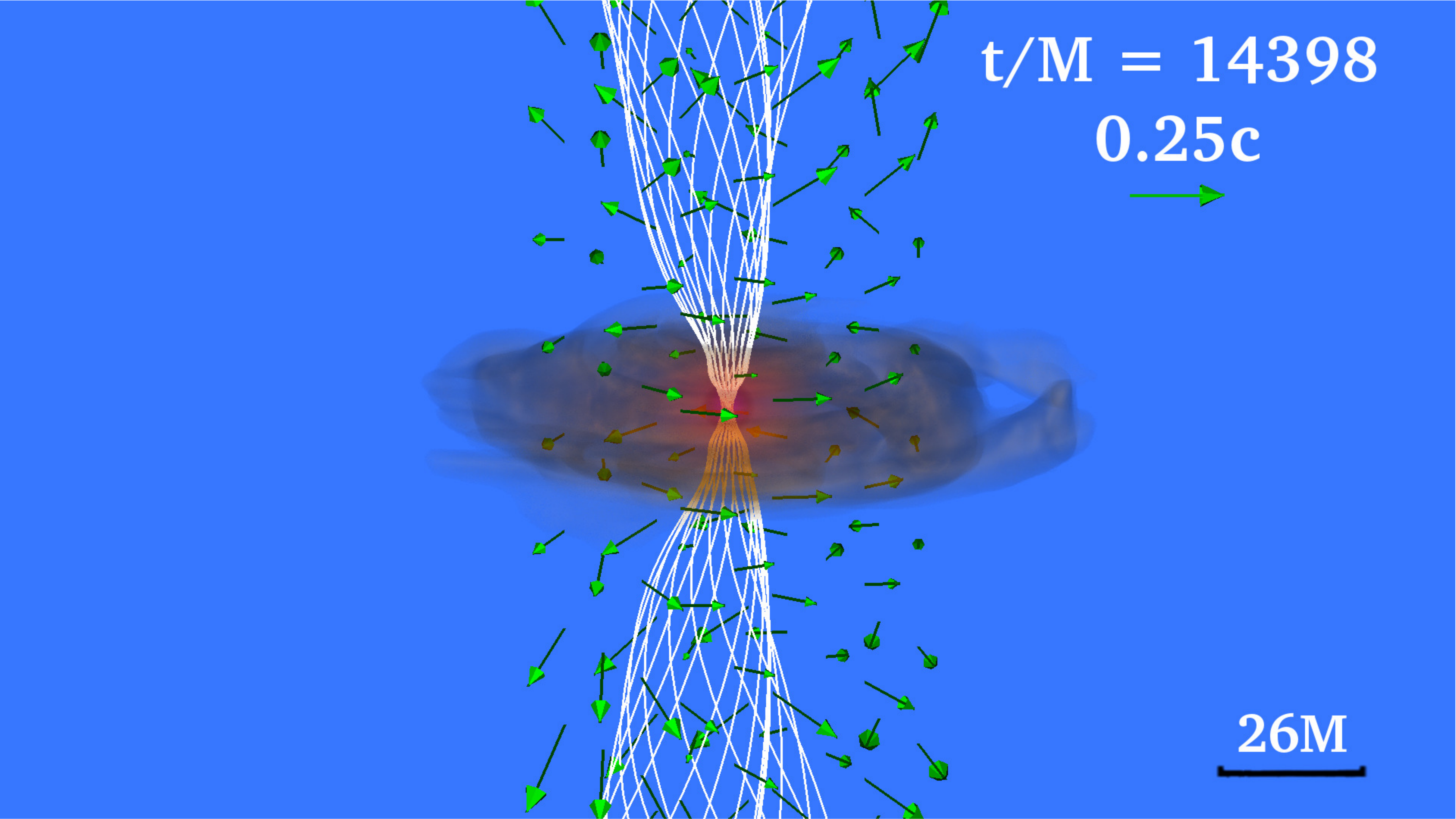}
  \includegraphics[width=5.9cm]{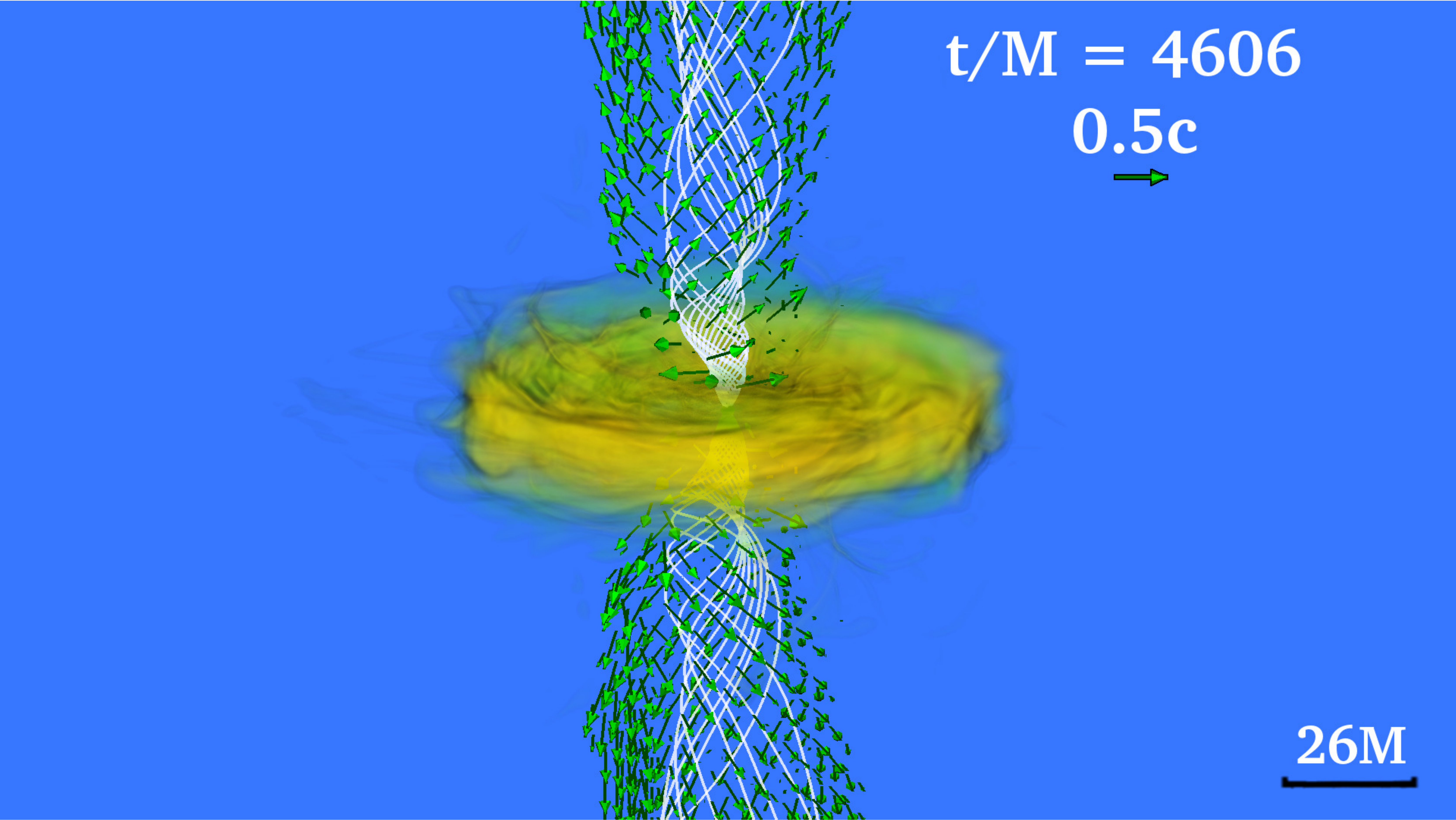}
  \includegraphics[width=5.9cm]{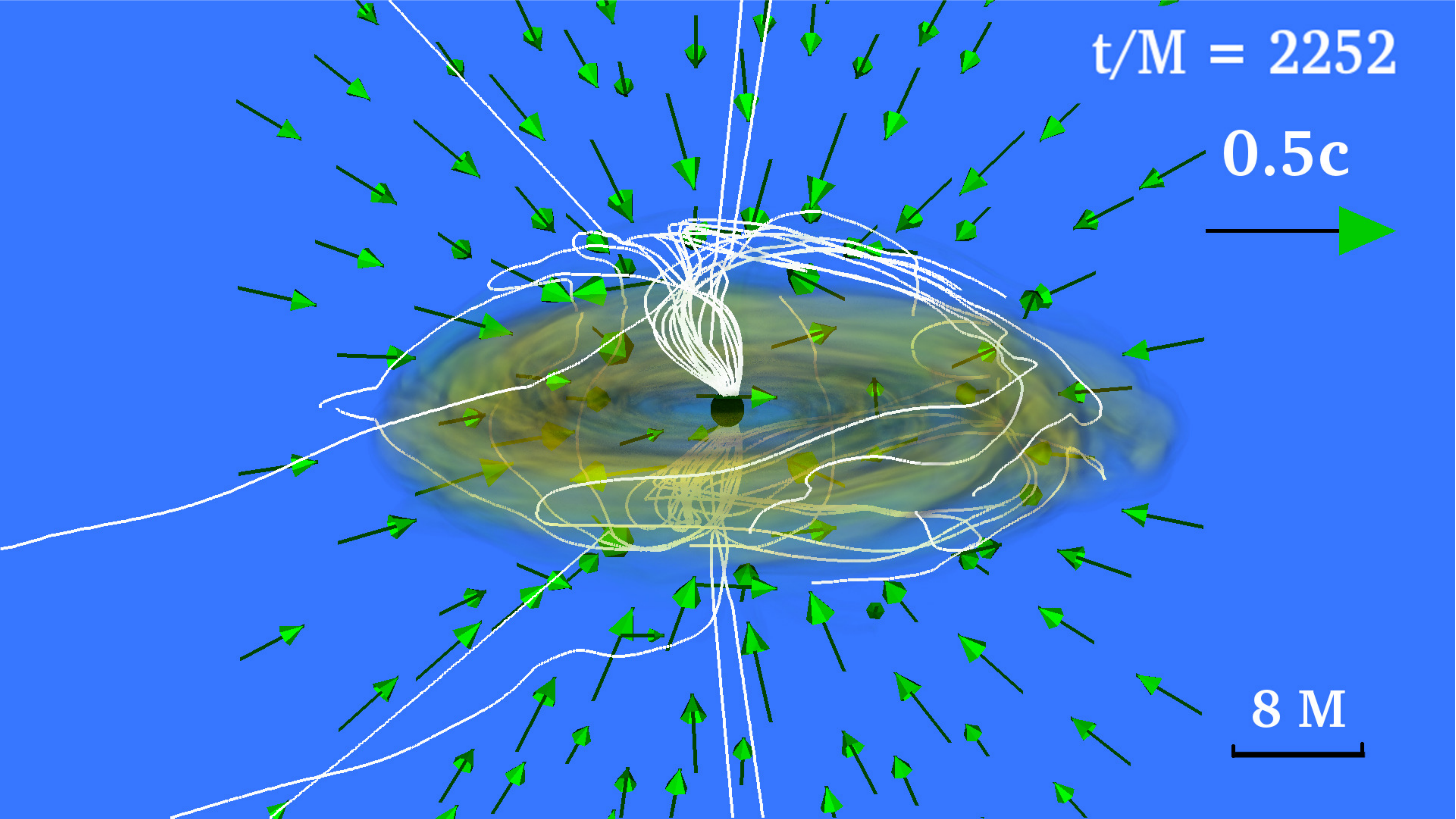}
  \includegraphics[width=5.9cm]{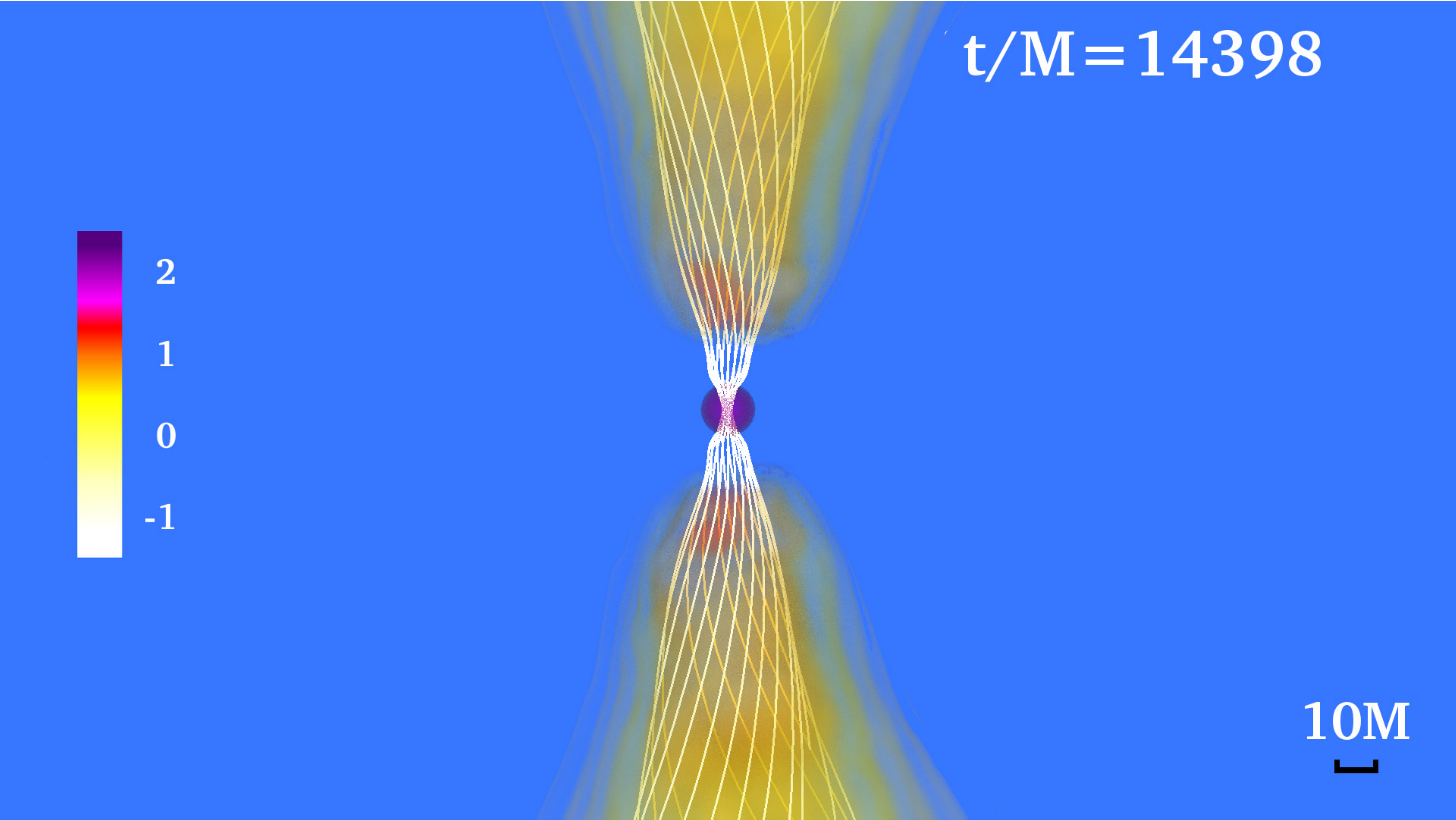}
  \includegraphics[width=5.9cm]{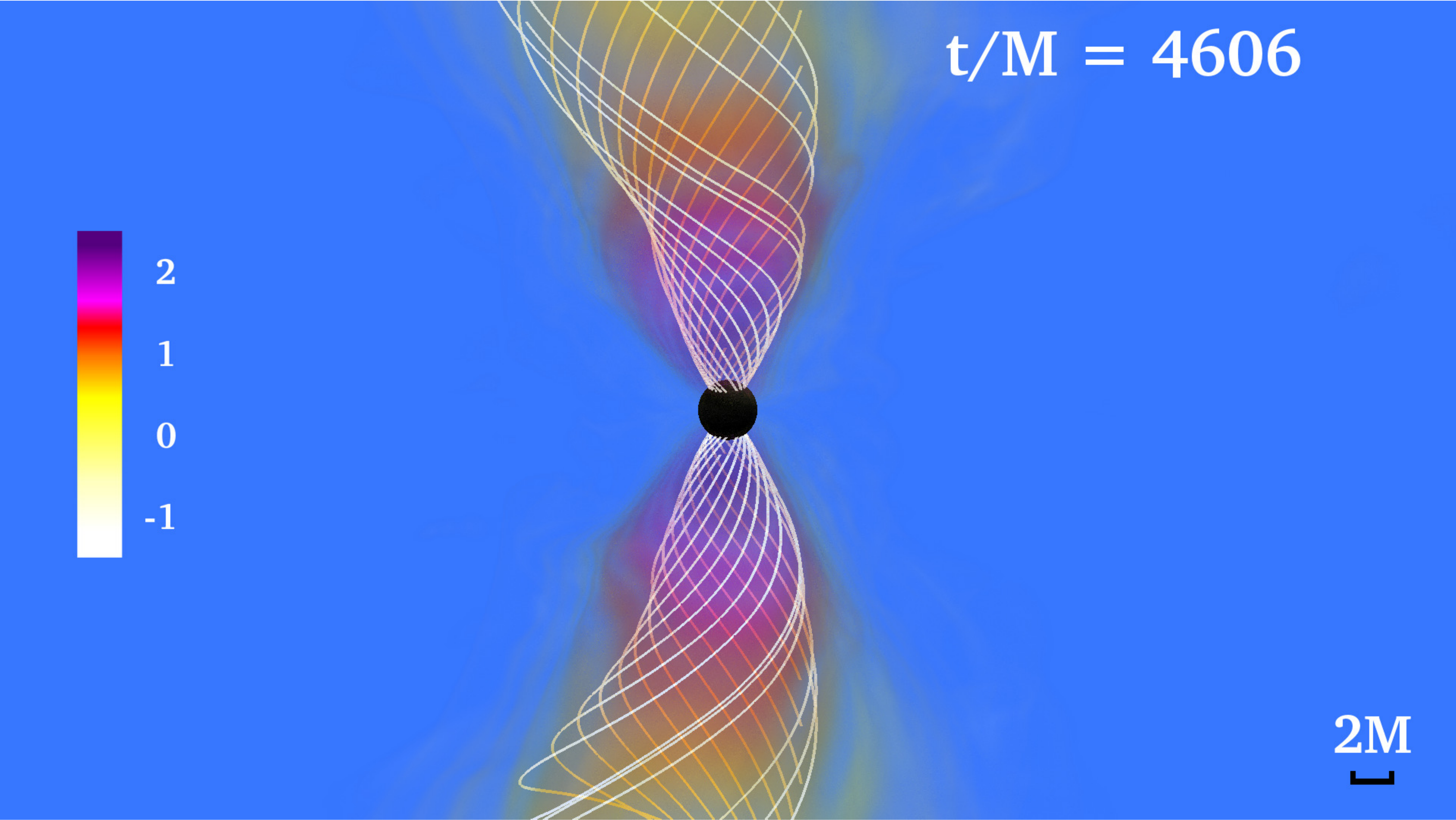}
  \includegraphics[width=5.9cm]{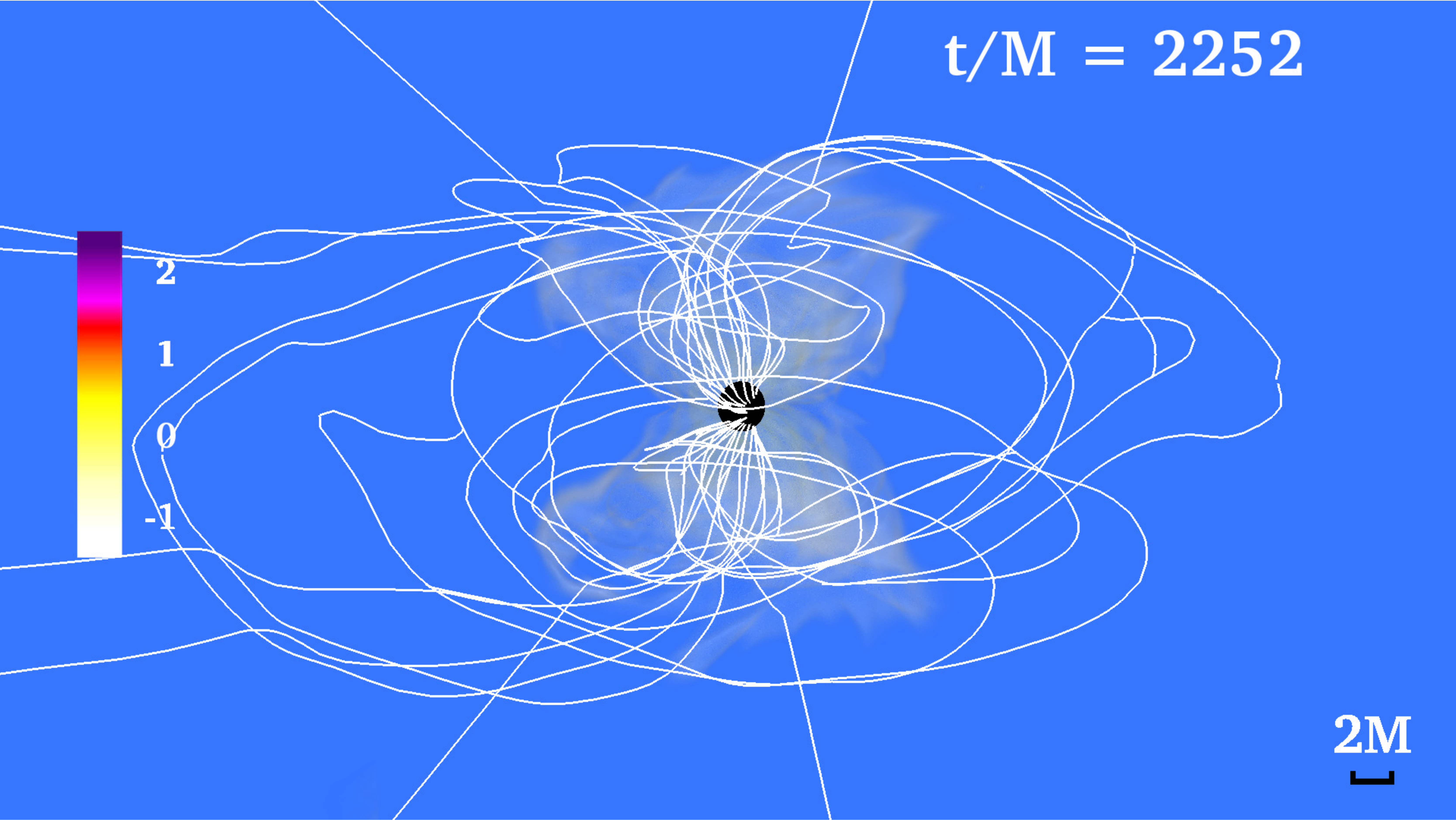}
  \caption{
    NS rest-mass density $\rho_0$ normalized to its initial maximum value (log scale) for
    a NSNS binary that forms: a stable, supramassive remnant (left column); a HMNS remnant that
    undergoes delayed collapse (middle column);  and a  remnant that undergoes prompt
    collapse (right column). Top row displays the NSs at the time of magnetic field insertion, while
   middle row displays the outcome once the  remnant has reached quasi-equilibrium. Bottom
row shows the force-free parameter $B^2/(8\pi\rho_0)$ (log scale). White lines represent
magnetic field lines, while arrows represent fluid velocity flow vectors.
The field lines form a tightly wound helical funnel and drive a jet following delayed collapse,
but not in the other two cases. Here $M=0.0136(M_{\rm tot}/2.74 M_\odot)\,{\rm ms} =
4.07(M_{\rm tot}/2.74 M_\odot)\,\rm km$; therefore quasi-equilibrium for the supramassive case 
(left column) is achieved at $t\sim 200$ ms [From~\cite{Ruiz:2017due}].
\label{fig:rmd}}
\end{figure}

In addition, \cite{Ruiz:2016rai} studied the impact of the magnetic configuration
on the jet launching time. For this the NSs were endowed with the pulsar-like interior + exterior
magnetic field generated
by the vector field in Eq.~\ref{eq:Aphi}. To reliable evolve the exterior magnetic field,
\cite{Ruiz:2016rai} adopted the atmosphere treatment previously used by~\cite{prs15}. As 
illustrated in Fig.~\ref{fig:BNS_ruiz},
a magnetically-driven jet is launched on the same time scale (see second column 
in~Fig.~\ref{fig:rmd}). Unlike in the BHNS case in~\cite{prs15}, where the magnetic field 
grows following BH formation, the MRI and magnetic winding in the HMNS already amplifies the magnetic
field  to saturation levels before the onset of collapse  to a BH. The incipient jet is then launched
by the BH + disk remnant due to the emptying of the funnel as matter accretes onto the BH, 
thereby driving the magnetic field regions above the BH poles to nearly force-free values
($B^2/8\pi\rho_0\gg 1$). Notice
that the initial magnetic field configuration affects the level of collimation of the incipient jet.
The opening half-angle of the pulsar-like magnetic field case is $\sim 25^\circ$, while for the
magnetic field confined to the stellar interior it is~$\sim 30^\circ$. The Lorentz factor 
in the outflow is $\Gamma_{\rm L}\sim 1.2$. Thus, the incipient jet is only mildly relativistic.
However, the force-free parameter inside the funnel is $B^2/8\pi\rho_0\sim 100$ (see bottom panel
of the second column in~Fig.~\ref{fig:rmd}), and therefore fluid elements can be accelerated
to $\Gamma_{\rm L}\sim 100$~\citep{Vlahakis2003}. The lifetime of the accretion disk
(jet's fuel) is $\sim 100(M_{\rm NS}/1.625M_\odot)\,\rm ms$ and hence consistent with
sGRB lifetimes~\citep{Bhat:2016odd,Lien:2016zny,Svinkin:2016fho,Ajello:2019zki}. The outgoing Poynting
luminosity is $L_{EM}\sim 10^{50.3}-10^{51.3}\,\rm erg/s$, roughly consistent with the luminosity
expected from the BZ effect~(see Eq.~\ref{eq:LBZ}) and the universal merger model~\citep{shapiro17}.
As this equation is strictly valid for highly force-free magnetospheres, it is likely that any deviation
from the expected Poynting luminosity is due to partial baryon-loaded surroundings.
%
%%%%%%%%%%%%%%%%%
%%% Ruiz BNS  %%%
%%%%%%%%%%%%%%%%%
%

To further assess the robustness of the emergence of the incipient jet in NSNS mergers, numerical
studies by~\cite{Ruiz:2019ezy,Ruiz:2020via} probed the impact of the NS spin and the orientation
of the seed poloidal magnetic field on the formation and lifetime of the HMNS, BH + disk remnant, and the
jet launching time. \cite{Ruiz:2019ezy} found that the larger the corotating NS spin, the more massive
the accretion disk, and hence the longer the jet's lifetime. In addition, the larger the NS
spin, the shorter the time delay between the peak GW and the emergence of the incipient jet.
On the other hand, the simulations of~\cite{Ruiz:2020via} suggest that there is a threshold
value of the inclination of magnetic dipole moment with respect to the orbital angular momentum $\vec{L}$
of the binary beyond which jet launching is suppressed. A  jet is launched whenever a net
poloidal magnetic flux with a consistent sign along $\vec{L}$ is accreted onto the BH once $B^2/8\pi
\rho_0 \gg 1$ above the BH poles. Tilted magnetic fields change the magnitude of this poloidal
field component. 

\begin{figure}
  \centering
  \includegraphics[width=18.3cm]{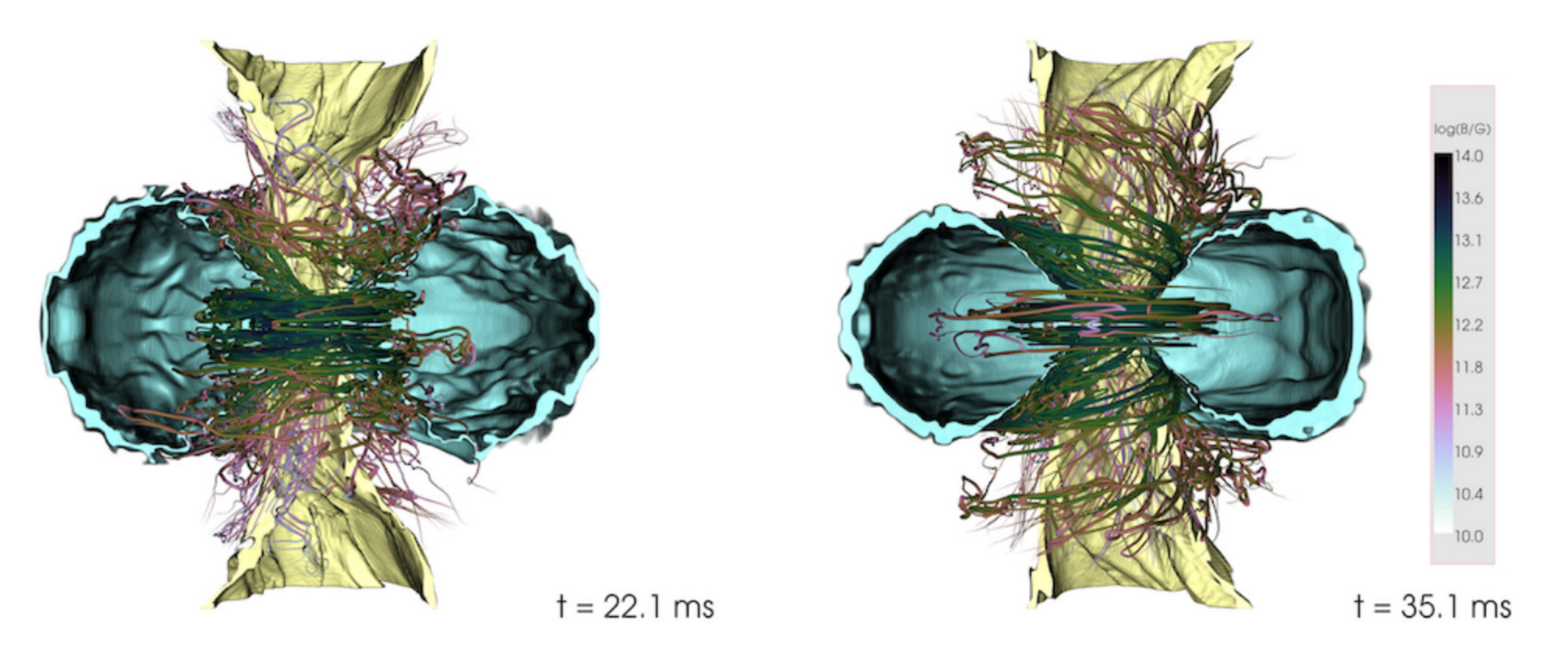}
  \caption{
    Magnetic field lines at $\sim 22\,\rm ms$ (left) and $\sim 32\,\rm ms$ (right)  following an NSNS merger,
    along  with two isosurfaces of rest-mass density $10^8$~(yellow) and $10^{10}\rm\,g/cm^3$~(cyan),
    cut off for~$y<0$ [From~\cite{Kawamura:2016nmk}].
\label{fig:burno_bns}}
\end{figure}

\cite{Kawamura:2016nmk} and~\cite{Ciolfi:2017uak} probed the effects of different EOSs, different mass
ratios, and different orientations of poloidal magnetic field confined to the NS interior, with strengths
$\sim 10^{12}-10^{15}\,\rm G$. The NSNS binaries were evolved with a resolution $\Delta x\gtrsim 177\,\rm
m$. These calculations found that after $22\,\rm ms$ following merger, an organized magnetic field
structure above the BH emerges, though  magnetically-driven outflow was not observed~(see Fig.~\ref{fig:burno_bns}).
The lack of an incipient jet is likely due to insufficient resolution to properly capture the
magnetic instabilities that boost the magnetic field strength to~$\gtrsim 10^{15.5}\rm G$, an
essential ingredient for jet launching, and/or to too short evolutions times. Notice that the ram-pressure
of the fall-back debris depends strongly on the EOS. More baryon-loaded surroundings require
stronger magnetic fields  to overcome the ram-pressure, delaying the launch of
the jet while the  fields amplify.

The previous numerical studies involved NSNS mergers leading to the formation of a transient HMNS
undergoing delayed collapse to a BH. The possibility of jet launching from a stable supramassive NS 
remnant has recently been investigated by~\cite{Ruiz:2017due}, \cite{Ciolfi2019} and \cite{Ciolfi:2020hgg}.
The calculation of~\cite{Ruiz:2017due} reported a long-term ($\sim 200\,\rm ms$) simulation of a supramassive NS
remnant initially threaded by a pulsar-like magnetic field.  It was found that magnetic winding induces
the formation of a tightly-wound-magnetic-field funnel within which some matter begins to flow outward
(see first column in Fig.~\ref{fig:rmd}). The maximum Lorentz factor in the outflow is $\Gamma_{\rm L}
\sim 1.03$, and the force-free parameter inside the funnel is $B^2/8\pi\rho_0\ll 1$.  The Poynting luminosity
is $\sim 10^{43}\rm\,erg/s$, and roughly matches the GR pulsar spindown luminosity~\citep{Ruiz:2014zta}.
These calculations suggest that a supramassive NS remnant probably cannot be the progenitor of a sGRB.
This has been confirmed by~the simulations of~\cite{Ciolfi2019} and~\cite{Ciolfi:2020hgg}, which
reported the emergence of an outflow with a maximum Lorentz factor of~$\Gamma_{\rm L}\lesssim 1.05$
after $\gtrsim 212\,\rm ms$ following the merger of a magnetized, low-mass NSNS. Recently, the
calculation of~\cite{Mosta:2020hlh}
suggested that neutrino effects may help reduce the baryon-load in the region above the poles of
the NS, inducing a growth of the force-free parameter in the funnel. They found a maximum Lorentz
factor of~$\Gamma_{\rm L}\lesssim 5.0$~inside the funnel. Thus, neutrinos processes may help
to trigger the launching of an incipient jet.
Finally, the numerical simulations of~\cite{Ruiz:2017inq}, who did not include neutrinos, probed whether or not
prompt collapse NSNS remnants (BHs with small accretion disks) can launch incipient jets. No evidence
of an outflow or magnetic field collimation was found~(see third column on Fig.~\ref{fig:rmd}). It was
argued that the KHI and MRI do not have enough time to amplify the magnetic field prior to BH formation,
and hence a jet can not be launched. 

Although supramassive NS or prompt collapse remnants may not launch magnetically-driven jets, they
may be the progenitors of fast radio bursts (FRBs) --a new class of radio transients
lasting less than a few tens of milliseconds~\citep{Lorimer:2007qn,Thornton:2013iua}.
\cite{Falcke:2013xpa} have suggested that magnetic field reconfigurations  during the collapse
of a supramassive NS can induce a burst of EM radiation consistent with that of typical FRBs.
\cite{Palenzuela:2013kra} studied EM counterparts from the inspiral and merger of a NSNS binaries
using full GR resistive MHD simulations. They found that the interaction between the stellar magnetospheres
extracts kinetic energy from the binary and powers radiative Poynting fluxes as large as $L_{\rm EM}\simeq
10^{41-44}(B/10^{12}G)^2\,\rm erg/s$ in a few milliseconds. Motivated by these results,
\cite{Paschalidis:2018tsa} performed numerical simulations  of prompt collapse NSNS mergers in
which the NSs are initially endowed with a pulsar-like magnetic field. Combining their numerical
results with population studies, they concluded that FRBs may be the
most likely EM counterpart of prompt collapse NSNSs, as previously claimed by~\cite{Totani:2013lia}.

\vspace{0.5cm}
%%%%%%%%%%%%%%%%%%%
%%%  GW170717   %%%
%%%%%%%%%%%%%%%%%%%
\subsection{\bf GW170817 and the NS maximum mass}
Event GW170817~\citep{TheLIGOScientific:2017qsa} marked not only the first direct detection
of a NSNS binary undergoing merger  via GWs but also the simultaneous detection of the
sGRB GRB 170817A, and kilonova AT 2017gfo, the latter with its afterglow radiation in the radio, optical/IR,
and X-ray bands~\citep{FERMI2017GCN,2017GCN.21517....1K}. These observations have been used
to impose constraints on the physical properties of a NS, and  in particular, on 
the maximum mass of a nonrotating spherical NS, $\maxtov$.
%
%%%%%%%%%%%%%%%%%%%
%%%  Bruno_BNS  %%%
%%%%%%%%%%%%%%%%%%%

\cite{2017arXiv171005938M} argued that following the merger of the NSNS progenitor of GW170817,
a transient HMNS is formed which collapses to a BH on a timescale of~$\sim 10-100\,{\rm ms}$, producing
the observed kilonova ejecta expanding at mildly relativistic velocities. This conclusion combined
with the GW observation, led to their tight prediction that~$\maxtov\lesssim 2.17M_\odot$ with $90\%$ confidence.
On the other hand,~\cite{Shibata2017}  summarized a number of their
relativistic hydrodynamic simulations favoring a long-lived, massive
NS remnant surrounded by a torus to support their inferred requirement of a strong neutrino
emitter that has a sufficiently high electron fraction ($Y_e\gtrsim 0.25$) to avoid an
enhancement of the ejecta opacity. This argument led then to the results that~$\maxtov\sim 2.15-2.25M_\odot$. A recently
review of these calculations by~\cite{Shibata2019}
using energy and angular momentum conservation laws  again lead to~$\maxtov\lesssim 2.3M_\odot$.
\cite{Rezzolla_2018} assumed that the transient GW170817 remnant
collapsed to a spinning BH once it had reached a mass close to but below the maximum mass of
a supramassive star. This assumption combined with their quasi-universal rotating NS model relations led to 
$\maxtov\lesssim 2.16^{+0.17}_{-0.15}M_\odot$. \cite{Ruiz:2017due} used the existence of the sGRB
GRB170817A, combined with their conclusion that  only a NSNS merger that forms an HMNS that undergoes
delayed collapse to a BH  can be the progenitor of an engine that powers an sGRB (see Fig.~\ref{fig:rmd}),
to impose the bound $\maxtov\lesssim 2.74/\beta$ (for low spin priors), where $\beta$ is s the ratio of the maximum
mass of an uniformly rotating NS (supramassive limit) to the maximum mass of a
nonrotating star. Causality arguments allow $\beta$ to be as high as $1.27$, while most realistic
candidate EOSs predict $\beta\simeq 1.2$, yielding  $\maxtov$ in the range $\sim 2.16-2.28M_\odot$.
If instead one assumes high spin priors in interpreting the data for GW170817 their maximum mass limit
becomes $\sim 2.22-2.35M_\odot$. Thus the different analyses seem to converge on a
value for~$\maxtov\sim 2.2-2.3M_\odot$.

%%%%%%%%%%%%%%%%%%%%
%%%  Erogstars   %%%
%%%%%%%%%%%%%%%%%%%%
%
\section{\bf Ergostars: potential multimessenger engines}
%\vspace{0.5cm}
In the previous two sections, we summarized GRMHD simulations  showing that the key requirement
for the emergence of a magnetically-driven jet is the existence of a spinning BH remnant surrounded
by an appreciable disk. In addition, these simulations also suggest that the BZ process is the
driving mechanism to power them.
\begin{figure}[h!]
  \begin{center}
    \includegraphics[width=8.5cm]{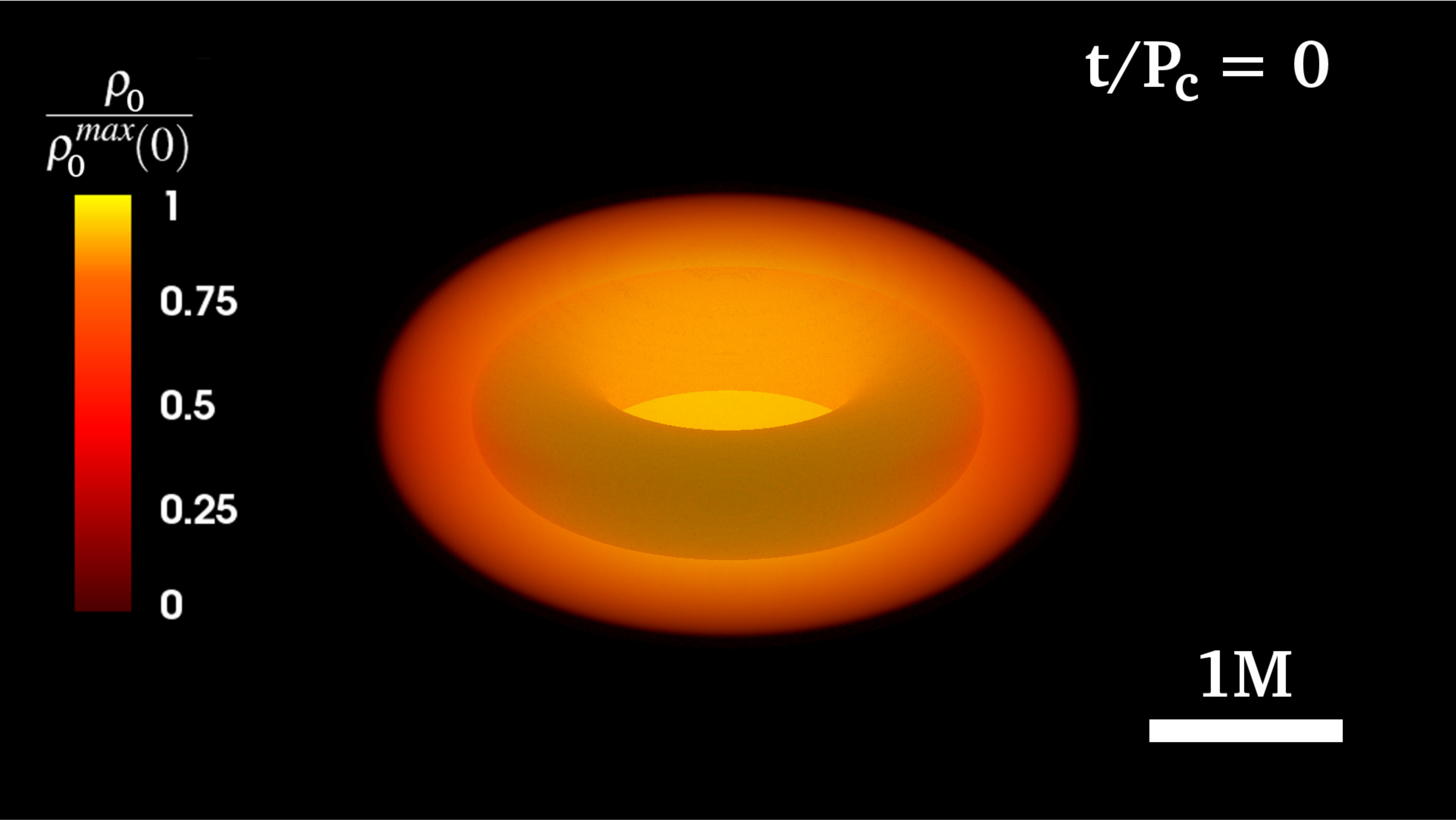}
    \includegraphics[width=8.5cm]{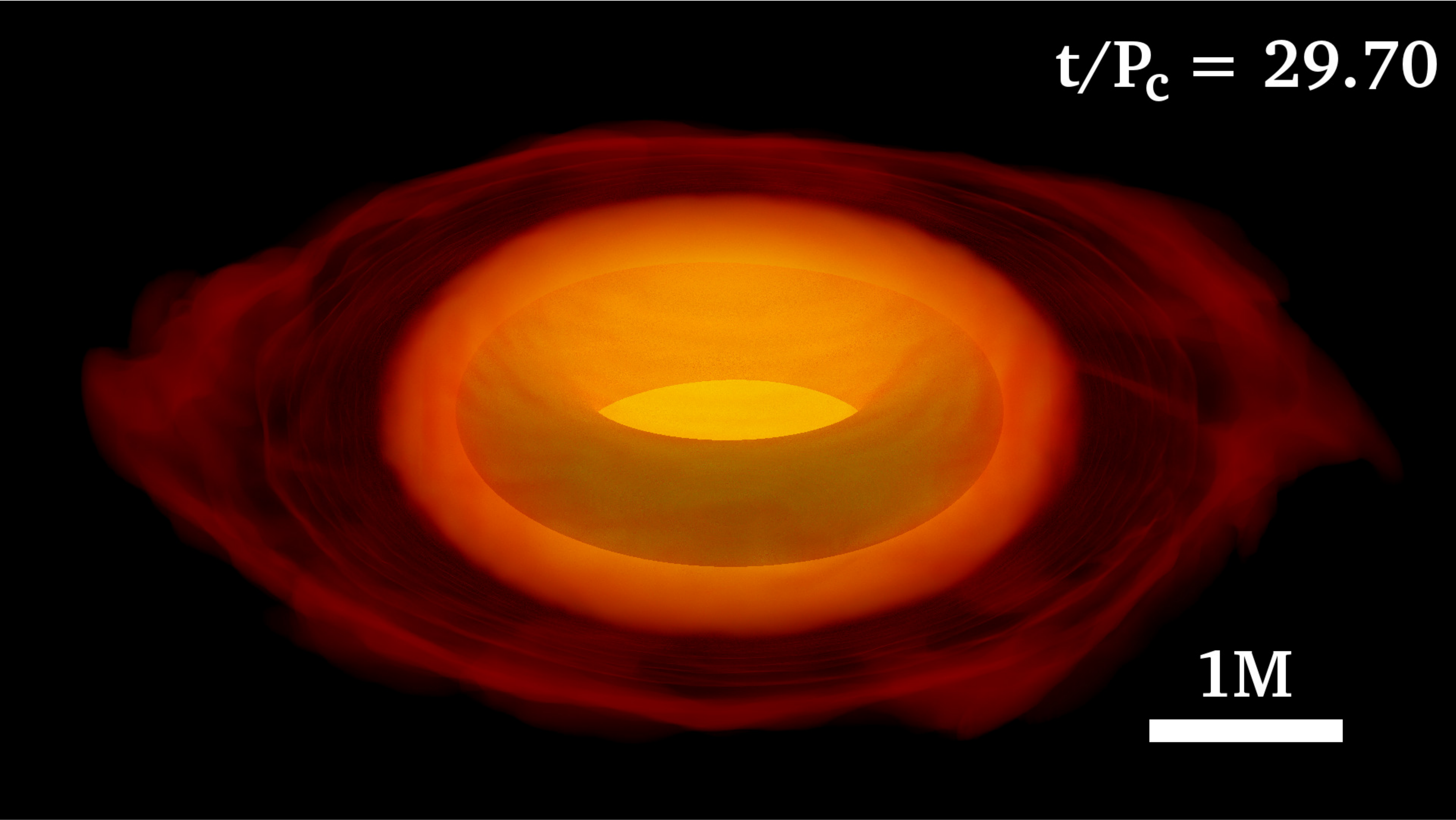}
    \end{center}
  \caption{Initial and final profiles of a dynamically stable ergostar modeled with the
    ALF2cc EOS (see Eq.~\ref{eq:eoscc}). The rest-mass density $\rho_0$ is normalized to its
    initial maximum value.
    The inner shaded torus indicates the position of the ergoregion. Here $P_{\rm{c}}$ is the initial
    rotation period measured at the point where the rest-mass density is maximum
    [From~\cite{Tsokaros:2019mlz}].}
\label{fig:5}
\end{figure}

The BZ  process can be explained using the membrane paradigm~\citep{Thorne86}, in
which the BH horizon is treated as a spherical, rotating conductor of finite resistivity.
The magnetic field lines  threading  the BH horizon transfer  rotational kinetic energy
from a spinning BH to an outgoing Poynting and matter flux. However, \cite{Komissarov:2002dj,
  Komissarov:2004ms,2005MNRAS.359..801K} has argued that the BH horizon is not the
``driving force'' behind the BZ  mechanism, but rather it is the ergoregion.
To disentangle the effects of the BH horizon and the ergoregion,~\cite{Ruiz:2012te}
performed force-free, numerical evolutions of magnetic fields on the {\it fixed} matter
+ metric background of an ``ergostar'' (a star with an internal ergoregion but no horizon)
modeled by the EOS of incompressible, homogeneous matter with constant total mass-energy density.
In addition, the same magnetic fields were evolved on the fixed background of a spinning BH.
\cite{Ruiz:2012te}~found that once the system reaches quasi-equilibrium, the configuration
of the EM fields and currents on both backgrounds are the same, in agreement with
\cite{Komissarov:2002dj,Komissarov:2004ms,2005MNRAS.359..801K}. These preliminary results
suggest that the BZ process is a mechanism driven by the ergoregion, and not by the BH horizon.

Recently,~\cite{Tsokaros:2019mlz,Tsokaros:2020qju} constructed the first dynamically
stable ergostars using compressible, causal EOSs based on the ALF2 and
SLy EOSs, but with their inner core replaced by the maximally
stiff EOS in Eq.~\ref{eq:eoscc}. The solutions are highly differentially rotating
HMNSs with a corresponding spherical compaction of $\mathcal{C}=0.3$. In principle, such
objects may form during
NSNS mergers.  Their stability was  demonstrated by evolving them in full GR for over
a hundred dynamical times ($\gtrsim 30$ rotational periods) and observing their
quasi-stationary behavior (see~Fig.~\ref{fig:5}). This stability was in contrast to earlier
$\Gamma=3$ polytropic models \citep{1989MNRAS.239..153K}, which proved radially unstable to collapse
\citep{Tsokaros:2019mlz}.

\begin{figure}
\begin{center}
\includegraphics[width=5.8cm]{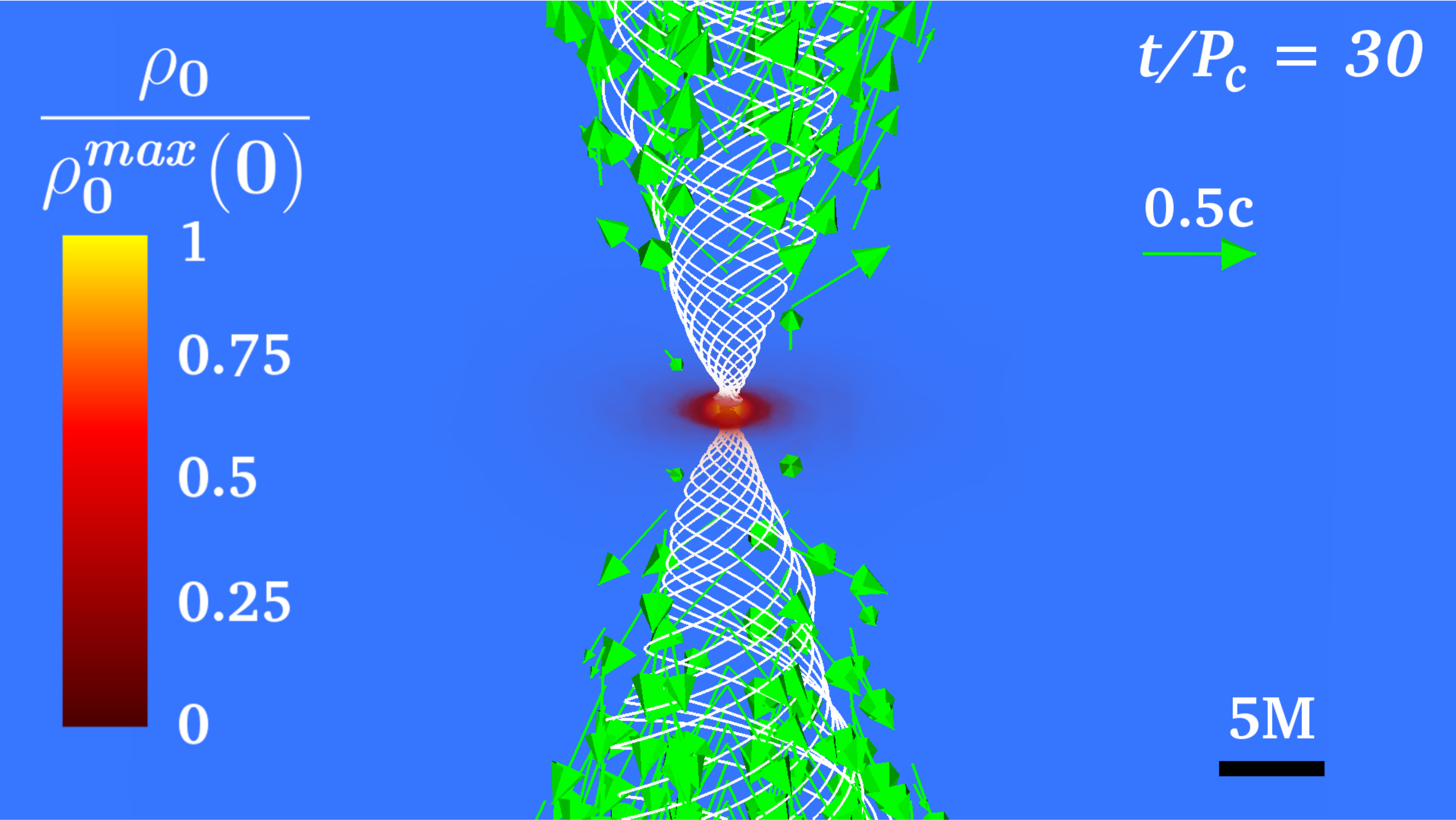}
\includegraphics[width=5.8cm]{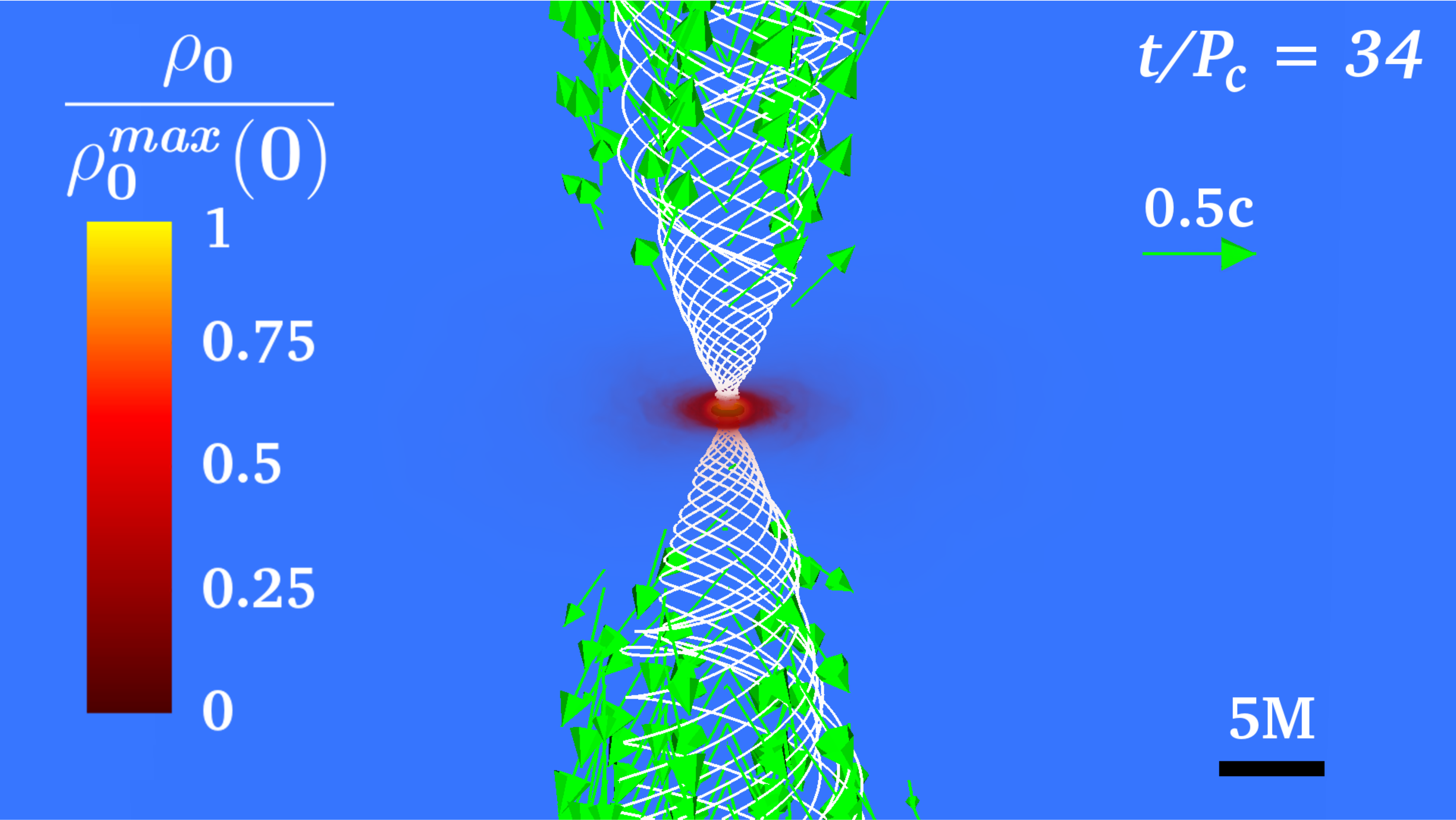}
\includegraphics[width=5.8cm]{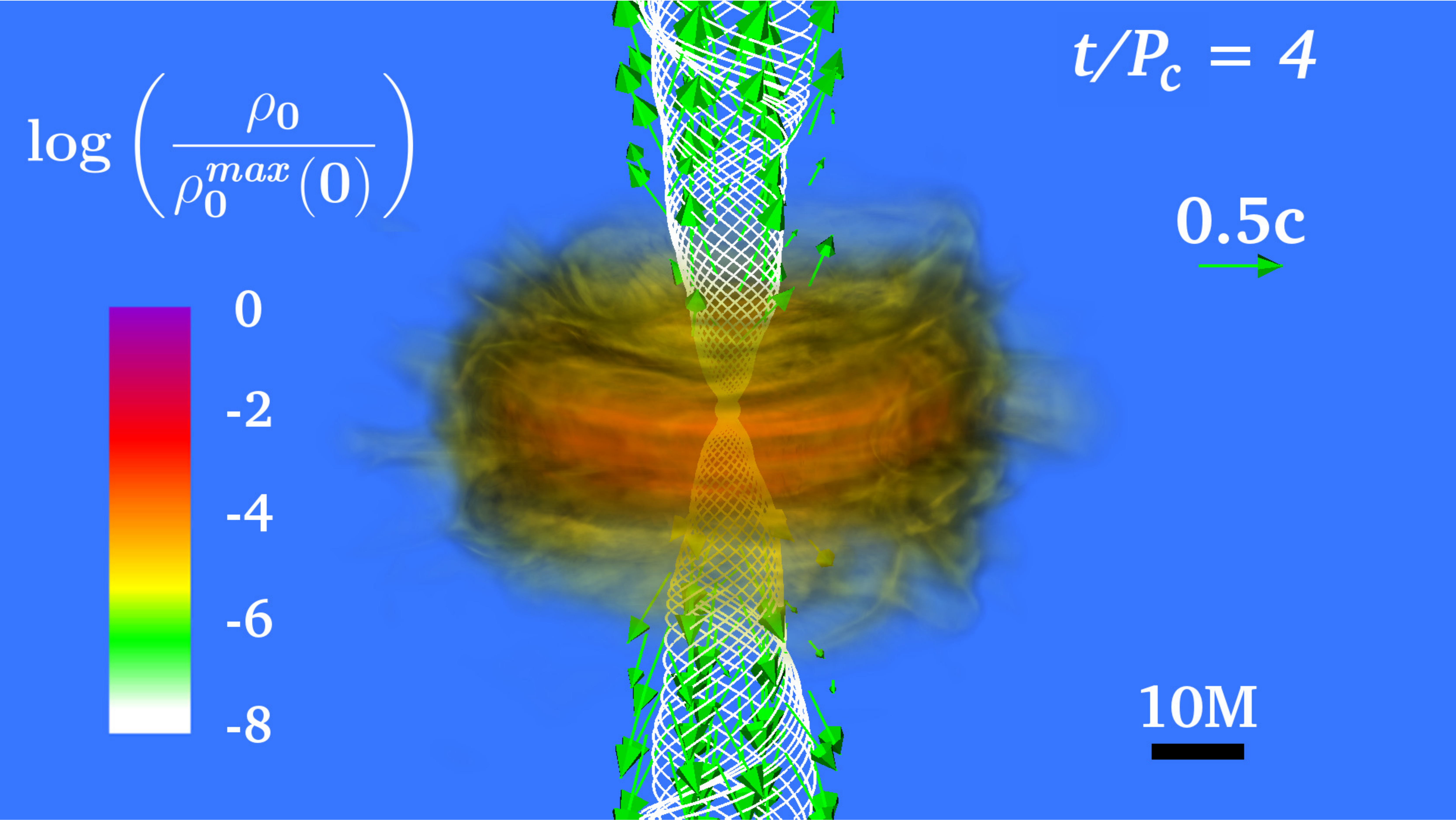}
\includegraphics[width=5.8cm]{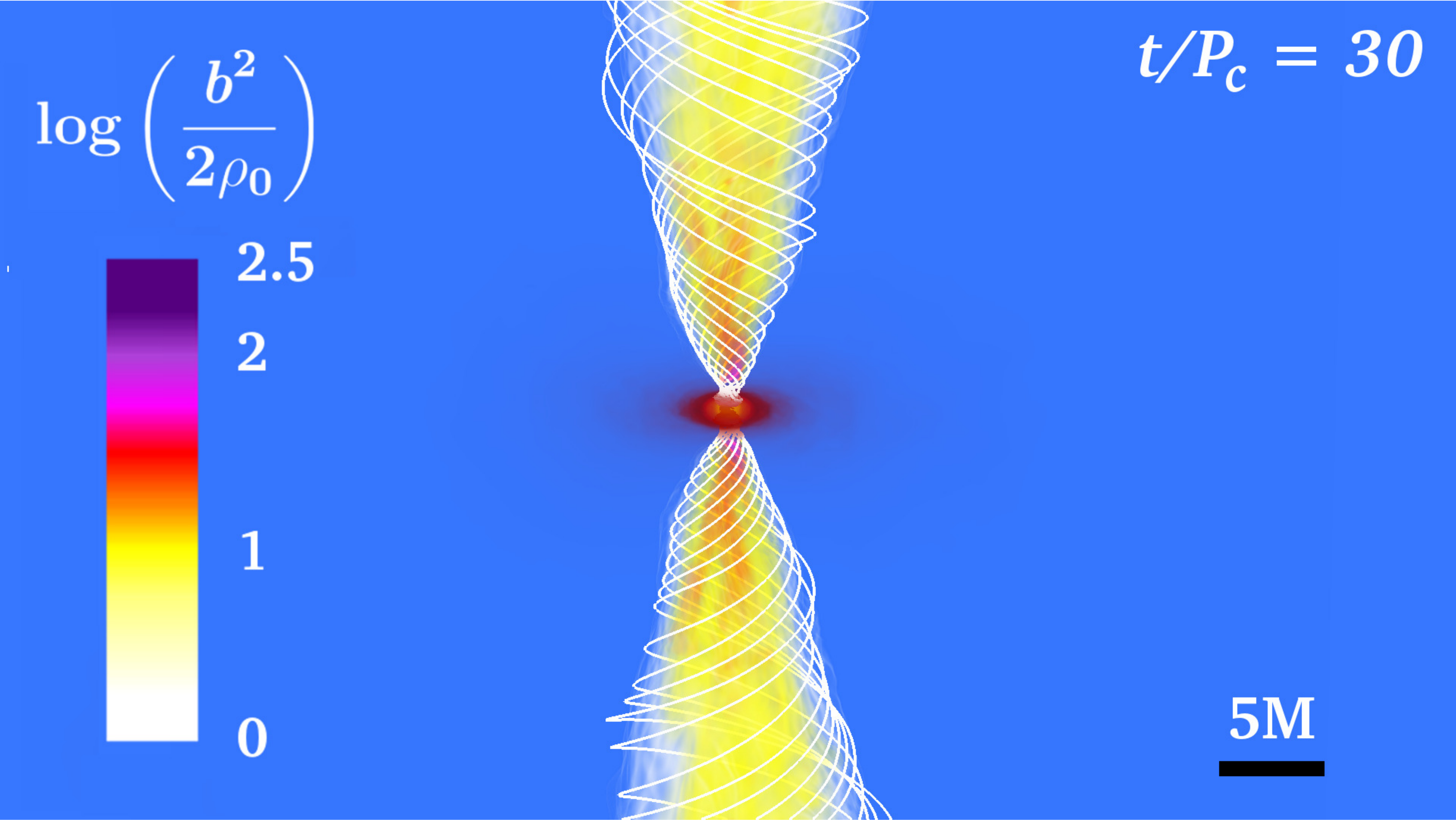}
\includegraphics[width=5.8cm]{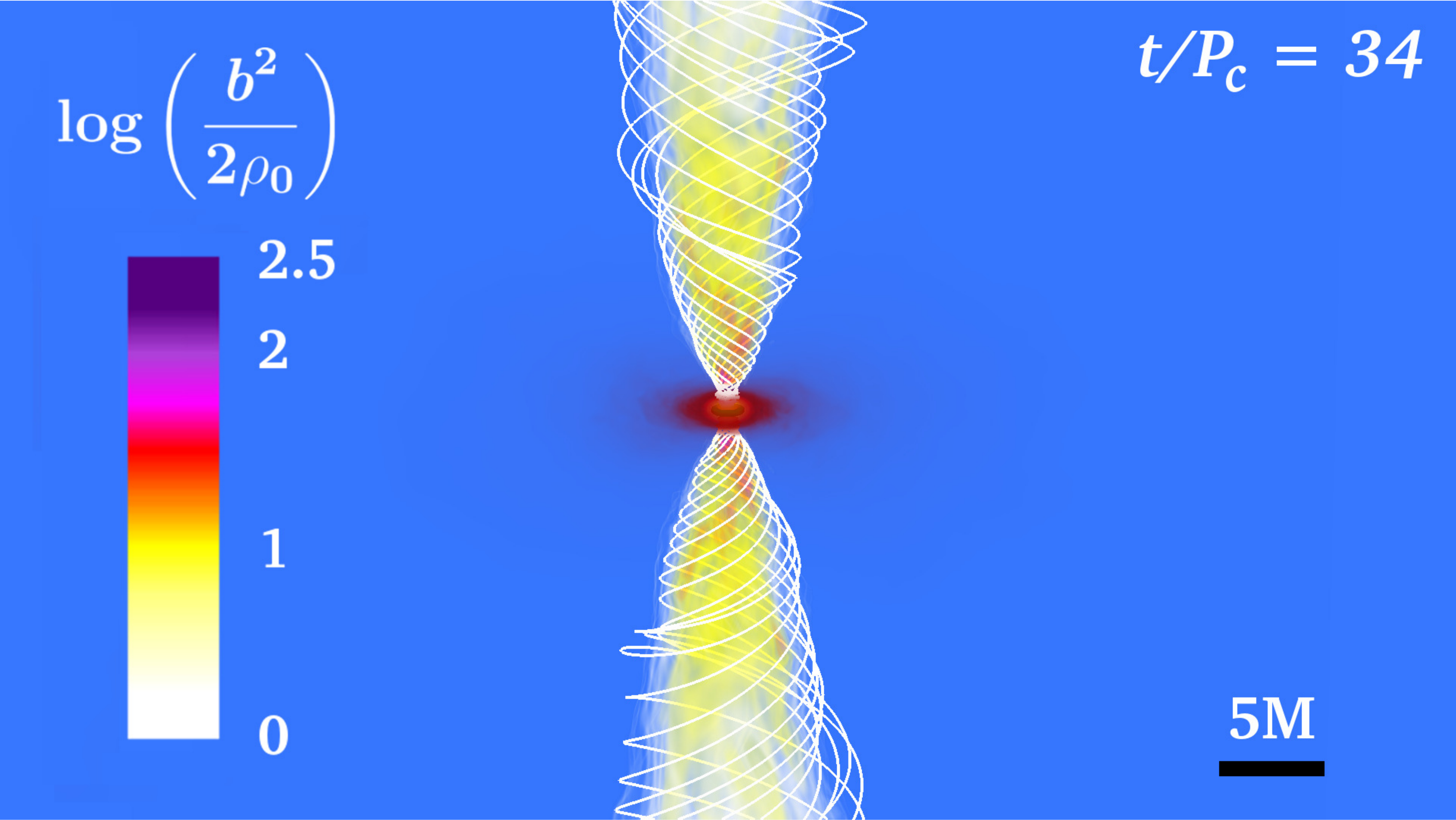}
\includegraphics[width=5.8cm]{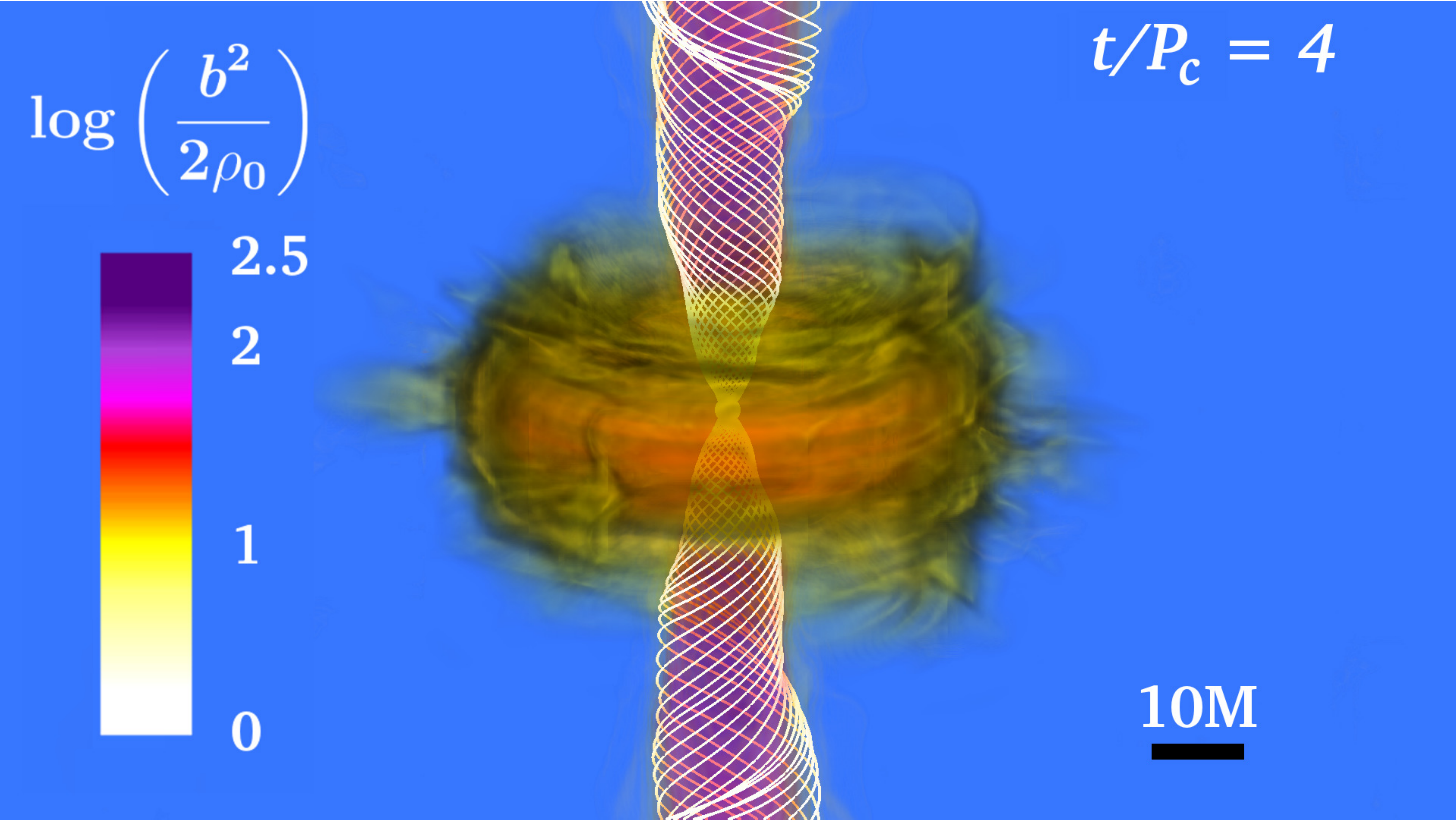}
\caption{Final profiles of the rest-mass density $\rho_0$ normalized to the initial maximum density (top),
  and the force-free parameter inside the helical magnetic funnel (bottom) for a standard HMNS (left), an ergostar
  (middle row), and BH + disk (right). White lines depict the magnetic field lines, while
  the arrows display fluid velocities. $P_c$ is the rotation period measure at the point where
  the rest-mass density is maximum. Here $M = 5.9\ \rm km$ and $b^2=B^2/4\,\pi$.
  [Fig.~1~from~\cite{Ruiz:2020zaz}].}
\label{fig:bfield}
\end{center}
\end{figure}

Using the above models, \cite{Ruiz:2020zaz} performed the first fully GRMHD simulations
of dynamically stable ergostars to assess the impact of ergoregions on launching
magnetically--driven outflows. In addition, and for comparison purposes, the evolution
of a standard magnetic HMNS without an ergoregion and a highly spinning BH surrounded by
a magnetized accretion disk were also considered.
The ergostar and the standard HMNS were initially endowed with a pulsar-like magnetic field
generated by the vector potential in~Eq.~\ref{eq:Aphi},
while the accretion disk was endowed with a poloidal magnetic field confined to the interior
(see Eq.~\ref{ini:Aphi_int}).
In all cases, after a few Alfv\'en times, the seed magnetic field is wound into a helical
structure from which matter begins to flow outward (see Fig.~\ref{fig:bfield}).
In the HMNS cases (ergostar and standard star), the maximum Lorentz factor in the outflow is
$\Gamma_{\rm L}\sim 2.5$, while in the BH + disk case~$\Gamma_{\rm L}\sim 1.3$. Therefore,
a mildly relativistic jet is launched  whether or not an ergoregion is present.
However, only in the BH + disk case does the force-free parameter reach $B^2/8\pi\rho_0
\gtrsim 100$, whereby the outflow can be accelerated to $\Gamma_L\gtrsim 100$ as required
by sGRB models~\citep{Zou2009}. These  simulations suggest that the BZ process
only operates  when a BH is present, though the Poynting luminosity in all cases is
comparable. Further studies are required to confirm this tentative conclusion.

%%%%%%%%%%%%%%%%%%%%%%
%   Acknowledgments
%%%%%%%%%%%%%%%%%%%%%%

\acknowledgements

We thank T. Baumgarte, C. Gammie, V. Paschalidis, and N. Yunes for useful discussions,
and members of the Illinois Relativity group undergraduate research team (K. Nelli, M. N.T
Nguyen, and S. Qunell) for assistance with some of the visualizations.
This work was supported by National Science Foundation Grant No. PHY-1662211 and
the National Aeronautics and Space Administration (NASA) Grant No. 80NSSC17K0070 to the
University of Illinois at Urbana-Champaign. This work made use of the
Extreme Science and Engineering Discovery Environment, which is supported by National
Science Foundation Grant No. TG-MCA99S008. This research is part of the Blue Waters
sustained-petascale computing project, which is supported by the National Science Foundation
(Grants No. OCI-0725070 and No. ACI-1238993) and the State of Illinois. Blue Waters
is a joint effort of the University of Illinois at Urbana-Champaign and its National Center
for Supercomputing Applications. Resources supporting this work were also provided by the
NASA High-End Computing Program through the NASA Advanced  Supercomputing Division at Ames
Research Center.

%%%%%%%%%%%%%%%%%%%%%%%%%%%%%%%%%%%%%%%%
\bibliographystyle{aasjournal}       %%%
\bibliography{references}            %%%
%%%%%%%%%%%%%%%%%%%%%%%%%%%%%%%%%%%%%%%%

\end{document}